\documentclass[aps,prd,preprint,superscriptaddress,showpacs,floatfix,nobibnotes,nofootinbib]{revtex4-1}

\usepackage{verbatim} 
\usepackage{latexsym}
\usepackage{amsmath}
\usepackage{amssymb}
\usepackage{graphicx}
\usepackage{longtable}
\usepackage[normalem]{ulem}
\usepackage[usenames, dvipsnames]{color}

\usepackage[colorlinks=true]{hyperref}
\usepackage{multirow}
\usepackage{dsfont}

\usepackage{setspace}

\newcommand{\bea}{\begin{eqnarray}}
\newcommand{\eea}{\end{eqnarray}}

\newcommand{\be}{\begin{equation}}
\newcommand{\ee}{\end{equation}}
\newcommand{\ba}{\begin{eqnarray}} 
\newcommand{\ea}{\end{eqnarray}} 
\newcommand{\nn}{\nonumber}
\numberwithin{equation}{section}

\DeclareMathAlphabet\mathbfcal{OMS}{cmsy}{b}{n} 

\begin{document} 

\title{Transport of hard probes through glasma}

\author{Margaret E. Carrington}
\affiliation{Department of Physics, Brandon University,
Brandon, Manitoba R7A 6A9, Canada}
\affiliation{Winnipeg Institute for Theoretical Physics, Winnipeg, Manitoba, Canada}

\author{Alina Czajka}
\affiliation{National Centre for Nuclear Research, ul. Pasteura 7,  PL-02-093  Warsaw, Poland}

\author{Stanis\l aw Mr\' owczy\' nski} 
\affiliation{National Centre for Nuclear Research, ul. Pasteura 7,  PL-02-093 Warsaw, Poland}
\affiliation{Institute of Physics, Jan Kochanowski University, ul. Uniwersytecka 7, PL-25-406 Kielce, Poland}

\date{June 1, 2022}

\begin{abstract}

We calculate the transverse momentum broadening $\hat q$ and collisional energy loss $dE/dx$ of hard probes traversing an evolving glasma during the earliest phase of a relativistic heavy-ion collision. We use a Fokker-Planck equation and apply a proper time expansion to describe the temporal evolution of the glasma. The correlators of the chromodynamic fields that determine the Fokker-Planck collision terms, which in turn provide $\hat q$ and $dE/dx$, are computed to fifth order. 
Both transport coefficients are strongly dependent on time. 
The maximum values they acquire before the proper time expansion breaks down are large:
$\hat q$ is of the order of a few ${\rm GeV^2/fm}$ and $dE/dx \sim 1~{\rm GeV/fm}$.
Their precise values depend on the probe's velocity ${\bf v}$,
the saturation momentum $Q_s$, and an IR regulator $m$ that is related to the confinement scale. 
We study the dependence of our results on these quantities. 
Different regularization procedures are analysed and shown to produce  similar results. 
We also discuss the validity of the proper time expansion and the compatibility of the approximations that are inherent in the derivation of the Fokker-Planck equation. 
We show that hard probes 
lose a comparable amount of energy when they propagate through the short-lived glasma phase, and  
the long-lasting hydrodynamic phase.
The conclusion is that the glasma plays an
important role in jet quenching.

\end{abstract}

\maketitle

\section{Introduction}
\label{intro}

Hard probes, due to their large momenta (or masses), are produced only through hard interactions with large momentum transfer at the earliest phase of a heavy-ion collision at the Relativistic Heavy Ion Collider (RHIC) and the Large Hadron Collider (LHC). The production mechanisms of heavy quarks and high $p_T$ partons are thus described by perturbative QCD. These particles propagate through the evolving medium probing QCD matter at different energy scales and different phases of the evolution of the system. During this propagation heavy quarks and high-$p_T$ partons lose a substantial fraction of their initial energy. 
Parton energy loss causes significant suppression of final high-$p_T$ hadrons, commonly known as jet quenching. The suppression of high-$p_T$ hadrons is treated as a signal of the formation of quark-gluon plasma, because only a deconfined state of matter could produce such significant braking of hard partons.  
The ultimate goal is to describe the full process of parton branching, jet structure and combined multi-stage interactions, see the review~\cite{Wiedemann:2009sh} and the references \cite{JETSCAPE:2021ehl,Kumar:2020vkx,Vujanovic:2020wuk} for recent progress. A simpler goal on which much effort has been focused is to understand the mechanisms of energy loss for an individual coloured probe in each phase.

The energy loss of a probe is caused by collisions and/or radiation and depends on the medium content and the dynamics of the system.
Collisional energy loss is similar for high-$p_T$ light partons and heavy quarks, but the situation with radiative energy loss is less clear. 
While there is no principal difference between the interactions of light and heavy particles with their surroundings, since they are both universally governed by QCD dynamics, 
the large masses of heavy quarks (charm and beauty)  make their evolution different when compared to high-$p_T$ light partons. Heavy quarks are particularly interesting because they are rare constituents of the quark-gluon plasma, and therefore may be thought of as external and clean probes of the medium. It is expected that the radiative energy loss of heavy flavours is noticeably reduced when compared to light flavours \cite{Dokshitzer:2001zm}. This phenomenon is called the ``dead-cone effect'', and is related to a restriction of the accessible phase space of radiated gluons. This argument leads to the conclusion that there should be a flavour hierarchy in radiative energy loss ordered by the masses of the quarks. When comparing quarks and gluons, there are also differences that are not related to kinematics. Different numerical factors are obtained, because quarks and gluons belong to different representations of the group SU$(N_c)$. Based on these arguments, it was expected that gluons would lose a bigger amount of energy than light quarks, and light quarks would lose more energy than heavy quarks. However, experimental findings on nuclear modification factors of charged hadrons do not confirm this expectation, see the review \cite{Dong:2019byy}. The conclusion is that the mechanism of parton energy loss is more complex than anticipated and not well understood to date. In this paper we consider primarily heavy quarks, but in some cases the corresponding results for high-$p_T$ light partons are also discussed. We study both collisional energy loss and the transverse momentum broadening coefficient $\hat q$ which, along with the emission probability of gluons, characterises radiative energy loss~\cite{Baier:1996sk}. 

The medium produced in a heavy ion collision quickly approaches equilibrium, within about\footnote{Throughout the paper we use the natural system of units with $c = \hbar = k_B =1$.} 1~fm. A long-lasting phase of equilibrated quark-gluon plasma follows, with lifetime approximately  10 fm. The system then hadronizes and passes through chemical and kinetic freezeouts. 
The spectra of heavy quarks are expected to be mostly shaped in the  equilibrium phase and for a long time the effect of the preceding non-equilibrium phases was entirely ignored. Only recently  progress has been made to quantify this influence. Below we will concisely describe the main characteristics of hard probes traversing the thermal QGP, which is relatively well understood. Next we will give a short overview of more recent work on non-equilibrated plasmas. To be specific, we discuss heavy quarks in these media, but similar reasoning holds for high $p_T$ light partons.

The Boltzmann equation provides a general method to describe a heavy quark embedded in a thermal medium, but it is difficult to solve. A heavy quark can be treated as a Brownian particle at both low and high $p_T$ for two reasons. First, the typical momentum exchange in the collision of a heavy quark with a plasma constituent is much smaller than the quark momentum itself. Second, the collisions are frequent and can be treated as uncorrelated with each other. 

The Boltzmann equation can be converted into the Fokker-Planck equation using the diffusion approximation \cite{Landau:10}, 
which is realized as follows. Because of the low concentration of heavy quarks in the plasma, their collisions with each other may be neglected, and only collisions with light quarks and gluons must be considered.  The small momentum exchange allows one to expand the transition matrix elements in the collision terms of the Boltzmann equation in powers of momentum transfer. One obtains the Fokker-Planck equation by performing momentum integrals over products of  approximated transition matrix elements and distribution functions of light quarks and gluons. The Fokker-Planck equation is usually much simpler to solve than the Boltzmann equation and it directly gives transport coefficients. Alternatively, the Brownian motion of heavy quarks can be studied via Langevin-type equations, see \cite{Moore:2004tg,Svetitsky:1987gq}.

Depending on the value of the transverse momentum, heavy quarks probe different features of the evolving matter. At low $p_T$, which is usually defined to be a few times larger than the temperature $T$ of the thermal bath, which varies between approximately 450 MeV and 150~MeV \cite{Shen:2011eg}, a heavy quark is a good probe to study thermalization and transport properties. In such a limit the quark's energy loss is dominated by collisional processes. Since the mass $m_Q$ of a heavy quark is large 
compared to the temperature of the medium, quarks need more time to adjust to their environment than light quarks or gluons. Thus their thermalization time is a factor of $m_Q/T$ longer than the thermalization time of light constituents \cite{Moore:2004tg}, see also \cite{Svetitsky:1987gq,vanHees:2004gq,Mustafa:2004dr}. At a medium-$p_T$ scale one is able to use heavy quarks as probes to study hadronization. At high $p_T$ (usually of the order of a few GeV or more) heavy quarks, light quarks and gluons all behave like hard probes, and their in-medium interactions are responsible for jet quenching. In this case, the major interest is in calculating the energy loss of the probe, which is expected to be radiation dominated. Various models have been developed using a transport equation in different forms to quantify the evolution, nuclear modification factors, and other properties of heavy quarks in a thermal medium, for a review see \cite{Prino:2016cni,Cao:2018ews}. 

When the plasma is in a non-equilibrium state, its properties are much more difficult to study. Two distinct pre-equilibrium phases can be identified: one just before the thermal quark-gluon plasma is formed, when the medium consists of quasi-particles with non-equilibrium distributions of momenta, and the strict earliest phase, when the medium is described in terms of strong classical gluon fields rather than partons.

If the system is made up of quasi-particles, the methods of kinetic theory can be used to study the dynamics of heavy quarks propagating through the medium. In Ref.~\cite{Das:2015aga} a method was developed to take into account the effect of this stage on the drag and diffusion coefficients of heavy quarks, and to compare the results with their respective equilibrium values. The coefficients are found from the Boltzmann equation as functions of the momenta of test partons, and have similar shapes and comparable sizes in kinetically equilibrated and out-of-equilibrium gluonic systems. The study was later developed to evaluate nuclear modification factors and elliptic flow coefficients at RHIC energies \cite{Das:2017dsh}. It is shown that when the pre-equilibrium phase with Kharzeev, Levin, Nardi (KLN) initial conditions for the transport equation is included in the evolution, the nuclear modification factor $R_{AA}$ can change by 20-25\% compared to the case where the initial stage is directly switched to hydrodynamic evolution. On the other hand, the pre-thermal phase does not have an important impact on the elliptic flow coefficient $v_2$. The dependence of the quenching and flow of heavy quarks on the parameters that define the pre-equilibrium evolution has 
also been 
studied in Ref.~\cite{Li:2020kax}. Such modelling shows significant uncertainties in $D$-meson suppression and flow at low $p_T$. Different configurations and various effects of the non-equilibrium dynamics have been shown to modify the charm quark $R_{AA}$ and $v_2$ \cite{Song:2019cqz} as well as the drag coefficient and the momentum broadening coefficient \cite{Song:2020tfm}. Since different non-equilibrium scenarios influence the calculations of observables, there are several aspects of these models that need to be understood before we can use them to interpret experimental data. 

The pre-equilibirum momentum distribution of quasi-particles is typically anisotropic, and in this case the weakly-coupled quark-gluon system is unstable due to colour plasma modes, see the review \cite{Mrowczynski:2016etf}. Chromodynamic fields are spontaneously generated, exponentially grow in time, and dominate the system's dynamics. Hard probes traversing such unstable systems have been studied in \cite{Carrington:2015xca,Carrington:2016mhd,Schenke:2008gg}. The energy loss and momentum broadening are strongly directionally dependent and rapidly grow as functions of time. Consequently, the magnitudes of $dE/dx$ and $\hat{q}$ can greatly exceed typical equilibrium values. The effect of an anisotropic QCD medium on heavy quarks was also studied recently in Ref.~\cite{Kumar:2021goi}, but the impact of instabilities was ignored. 
Strong dependence of various transport coefficients on the direction and strength of the momentum anisotropy of the QCD medium is observed. In Ref.~\cite{Hauksson:2021okc} a Keldysh-Schwinger approach was used to calculate the momentum broadening of an anisotropic plasma with the unstable modes eliminated. 
The collision kernel obtained for such a case is shown to lead to a significant decrease in $\hat q$ when compared to its isotropic counterpart, and features mild angular dependence.

The earliest phase of the collision is the glasma which, in the framework of a Color Glass Condensate (CGC) description, is made mostly of gluon fields that can be treated as classical long-wavelength fields generated by valence quarks acting as colour sources (see, for example, the review \cite{Gelis:2012ri}). 
The glasma state is populated with strong chromodynamic fields, and is characterized by large anisotropies and high energy density. These properties lead us to expect that the initial dynamics could significantly influence the propagation of high-$p_T$ particles. In Ref.~\cite{Aurenche:2012qk} a first attempt was made to assess the importance of 
synchrotron-like gluon emission from fast partons due to interaction with coherent glasma colour fields, and it was found that the effect is much smaller than the radiative energy loss from the quark-gluon plasma. However, this result was obtained using a small-angle approximation and may not give full information on the impact of this phase on hard partons. 

More recent studies have revisited the problem using different approaches. In Refs.~\cite{Mrowczynski:2017kso,Carrington:2020sww} the effect of the glasma on heavy quarks was studied through a properly formulated Fokker-Planck equation, and it was found to be large. This finding stimulated further interest in this direction. In Ref.~\cite{Pooja:2021tkd} the diffusion of heavy quarks in the evolving glasma was compared to their Brownian motion in a thermalized medium with the same energy density. 
In both systems, the average transverse momentum broadening was shown to significantly depend on the value of the saturation momentum $Q_s$. Results were comparable in the two approaches for small $Q_s$. 
Calculations in both the weak and dense limits of the glasma have been done using real-time lattice simulations \cite{Ipp:2020mjc,Ipp:2020nfu}. The results of these calculations show that the parameter $\hat q$ depends strongly on time and the orientation of the probe's momentum. One can also study the behaviour of heavy quarks influenced by gluon fields using the Wong equations, which is an alternative to the Fokker-Planck equation that does not use the diffusion approximation. This method was applied in Refs.~\cite{Ruggieri:2018rzi,Liu:2019lac,Sun:2019fud} to emphasize the remarkable impact of the evolving glasma on diffusion, medium modification and the flow of heavy quarks. It was also used in Ref.~\cite{Liu:2020cpj} to study the energy loss of heavy quarks including the effect of back-reaction. The dynamics of the earliest phase can also be  explored using classical-statistical simulations, and this approach was used to compute the momentum diffusion coefficient of a heavy quark in Ref.~\cite{Boguslavski:2020tqz}.


The goal of the current work is to provide an extension of the analysis of the transport of heavy quarks propagating through glasma which was proposed in \cite{Mrowczynski:2017kso}, and further developed in \cite{Carrington:2020sww}. The most important results of the current work were recently presented in the letter~\cite{Carrington:2021dvw}.
We use a Fokker-Planck equation whose collision terms encode information about
glasma dynamics through correlators of strong chromoelectric and chromomagnetic fields\footnote{For convenience, we neglect henceforth the prefix ``chromo'' when referring to chromoelectric or chromomagnetic fields. Since we study QCD only, this should not cause confusion.}.
We calculate the relevant correlators using a CGC approach combined with an expansion in the proper time $\tau$. This analytical approach to solve the classical Yang-Mills equations was originally proposed in~\cite{Fries:2006pv}, and developed further in~\cite{Chen:2015wia}. 
The key point is that because of the short lifetime of the glasma, the proper time can be treated as an expansion parameter (the dimensionless small parameter is $\tau Q_s$, where $Q_s$ is the saturation scale). 
Solutions for the glasma fields, and the associated electric and magnetic fields,
can be found to any order in~$\tau$ in terms of the initial gauge potentials using a recursive method. 
In our previous work \cite{Carrington:2020sww} we included terms up to first order in the~$\tau$ expansion to develop our method and explore its general features. In this paper we include terms up to fifth order in~$\tau$, which will allow us to study the radius of convergence of the expansion, and to obtain reliable results. We will also present a careful analysis of several technical aspects of our method which were not explored in~\cite{Carrington:2020sww}. We study the dependence of our results on the choices of the numerical values of two parameters that are a necessary component of the CGC approach: the saturation scale and an infra-red regulator that is related to the QCD confinement scale. In addition, our calculation requires the choice of a regularization method to tame an ultra-violet singularity. 
We will discuss different methods of  regularization and demonstrate that our results depend only weakly on 
the form that is used. 
We will present some comparisons between the results obtained for heavy quarks and high $p_T$ light partons.

Finally we add that the work described in this article  is a part of a bigger project, which aims at exploring the general properties of the glasma at very early times. Our previous papers~\cite{Carrington:2020ssh,Carrington:2021qvi}, in which we calculated the energy-momentum tensor and various observables that can be obtained from it, are also part of this main project. The proper time expansion is used in all of these calculations, and our analyses provide valuable information about the extent to which this method can be applied to glasma calculations. 

This paper is organised as follows. In Sec.~\ref{FP-eq} we describe the  Fokker-Planck equation, with particular emphasis on its collision terms and their relationship to two transport coefficients of hard probes in a glasma: collisional energy loss and momentum broadening. In Sec.~\ref{CGC-all} we explain how to calculate correlators of  chromodynamic fields, which determine the transport coefficients we are interested in. In Sec.~\ref{CGC} we review the basics of the McLerran-Venugopalan (MV) model \cite{McLerran:1993ni,Kovner:1995ts,Kovner:1995ja}, which provides the forms of the pre-collision potentials, and gives the boundary conditions that connect pre-collision and post-collision glasma potentials. In Sec.~\ref{sec-prop-time} we describe how we use a proper time expansion to represent glasma potentials and chromodynamic fields in the post-collision region. The correlators of the initial gauge potentials, which are the basic building blocks of the whole methodology, are discussed in Sec.~\ref{Before-LC-corr}, and in Sec.~\ref{field-corr} the electric and magnetic field correlators are derived. Section~\ref{sec-results} is devoted to a detailed discussion of our results. In Sec.~\ref{sec-res-0} we consider the different variables and parameters that influence the results obtained for collisional energy loss and momentum broadening. In Sec.~\ref{sec-res-a} we discuss the time evolution of these transport coefficients. We also discuss several technical aspects of the calculation. In Sec.~\ref{sec-res-b} we study the dependence of our results on the probe's velocity and its initial space-time rapidity. The dependence on UV and IR energy scales is studied in Sec.~\ref{sec-res-d}, and in Sec.~\ref{sec-res-c} we discuss our regularization method and demonstrate that results depend only weakly on the method that we choose. The limitations of our approach are analysed in Sec.~\ref{sec-res-e}. The impact of the glasma on jet quenching is evaluated in Sec.~\ref{sec-res-add}. We summarize our results and make some concluding remarks in Sec.~\ref{sec-conclusions}.

\section{Fokker-Planck equation}
\label{FP-eq}

The Fokker-Planck equation has been frequently employed to study the transport of heavy quarks across a thermalized quark-gluon plasma, see, for example \cite{Moore:2004tg,Svetitsky:1987gq,vanHees:2004gq,Mustafa:2004dr}. In this paper our aim is to study the transport of both heavy quarks and high-$p_\perp$ light partons through glasma in the earliest period of its temporal evolution. More specifically, we focus on the situation where the hard probes interact with the soft classical gluon fields of the glasma, and not with quasi-particles, which emerge at later stages. 

The formulation of the method and the derivation of the Fokker-Planck equation that we use is presented in Ref.~\cite{Mrowczynski:2017kso}. For the convenience of the reader we review the main points below. We comment that although the derivation was originally presented in the context of heavy quarks traversing a glasma, the formalism can also be used to study relativistic light partons, as long as the diffusion approximation is applicable.
For heavy quarks the method can be used for a broad range of velocities, and therefore provides much richer information about their spectra than is the case for light high-energy partons.

When a heavy quark is embedded in a glasma it is subject to stochastic processes due to the action of colour forces. The corresponding distribution function $Q(t,{\bf r},{\bf p})$ can  therefore be decomposed into regular and fluctuating components as follows\footnote{We denote three-vectors  as ${\bf x}=(x^1,x^2,x^3)$, and they are indexed by $\alpha,\beta \in (1,2,3)$.}
\be
\label{reg-fluc}
Q(t,{\bf x},{\bf p}) = \langle Q(t,{\bf x},{\bf p}) \rangle 
+ \delta Q(t,{\bf x},{\bf p}) ,
\ee
where $t$ is time, ${\bf x}$ is position, ${\bf p}$ is momentum and $\langle \cdots \rangle$ denotes a statistical ensemble average over events in a relativistic heavy ion collision. The regular contribution, denoted $\langle Q(t,{\bf x},{\bf p}) \rangle$, is assumed to be colour neutral and gauge independent, and is expressed as   
\be
\label{whiteness}
\langle Q(t,{\bf x},{\bf p}) \rangle  = 
n (t,{\bf x},{\bf p}) \, \mathds{1},  
\ee
where $\mathds{1}$ is a unit matrix in colour space. We use $\delta Q(t,{\bf x},{\bf p})$ to denote the fluctuating part and we assume that $\langle \delta Q \rangle =0$. It is also assumed that the regular part is a slowly varying function of time and space and is much larger than the fluctuating part. With these conditions, starting from a Vlasov-type equation, one is able to obtain the transport equation in the Fokker-Planck form~\cite{Mrowczynski:2017kso}, which reads
\be
\label{F-K-eq}
\Big({\cal D} - \nabla_p^\alpha  X^{\alpha\beta}({\bf v}) \nabla_p^\beta - \nabla_p^\alpha  Y^\alpha ({\bf v}) \Big) n(t, {\bf x},{\bf p}) = 0,
\ee
where ${\bf v}={\bf p}/E_{\bf p}$ is the velocity of the quark with $E_{\bf p}=\sqrt{{\bf p}^2 + m^2_Q}$. We also use ${\cal D} \equiv  \frac{\partial}{\partial t} + {\bf v}\cdot \nabla $ for the substantial, or material, derivative. The collision terms entering the Fokker-Planck equation~(\ref{F-K-eq}) are given by
\be
\label{Y-defi}
Y^\alpha({\bf v}) \, n({\bf p}) = \frac{1}{N_c} {\rm Tr} \big[ \big\langle \mathcal{F}^\alpha(t, {\bf x}) 
\delta Q_0({\bf x}-{\bf v} t,{\bf p})\big\rangle \big],
\ee
and
\be
\label{X-defi}
X^{\alpha\beta}({\bf v}) \equiv
\frac{1}{2N_c} \int_0^t dt' \: {\rm Tr}\big[ \big\langle \mathcal{F}^\alpha(t, {\bf x}) 
\mathcal{F}^\beta \big(t-t', {\bf x}-{\bf v} t'\big) \big\rangle \big] ,
\ee 
where $\delta Q_0 \equiv \delta Q(t=0,{\bf x},{\bf p})$ is the initial condition. The Lorentz colour force entering the collision terms (\ref{Y-defi}) and (\ref{X-defi}) is $\mathbfcal{F} (t,{\bf x}) \equiv g \big({\bf E}(t,{\bf x}) + {\bf v} \times {\bf B}(t,{\bf x})\big)$, where $g$ is the coupling constant. The electric ${\bf E}(t,{\bf x})$ and magnetic ${\bf B}(t,{\bf x})$ fields are given in the fundamental representation of the SU($N_c$) group.\footnote{The generators of the SU($N_c$) group are defined through $[ t^a,t^b ] =  i f^{abc}t^c$, where $f^{abc}$ are the structure constants, and 
${\rm{Tr}}(t^a t^b)=\delta^{ab}/2$.} 

In equilibrium, the Boltzmann distribution function $n^{\rm eq}({\bf p}) \sim \exp (-E_{\bf p}/T) $, where $T$ is the temperature of the system, should solve the Fokker-Planck equation. This requires a relation between $X^{\alpha\beta} ({\bf v})$ and $Y^\alpha({\bf v})$ of the form
\ba
\label{XY}
Y^\alpha({\bf v}) = \frac{v^\beta}{T} X^{\alpha\beta} ({\bf v}),
\ea
where $T$ is the temperature of an equilibrated quark-gluon plasma that has the same energy density as the glasma, or equivalently the temperature the glasma would have, if it equilibrated without expanding.
Our calculation gives no information about this temperature, but
in Sec.~\ref{sec-res-0} we discuss how to estimate its value. 
Since the formula (\ref{Y-defi}) is difficult to apply to a non-equilibrium system, we use the relation~(\ref{XY}) to determine $Y^\alpha({\bf v})$. As we will show below, the quantity $Y^{\alpha}({\bf v})$ is needed to obtain $dE/dx$, but it is not required for a calculation of $\hat q$.

When the system under consideration is translationally invariant, the tensor $X^{\alpha\beta} ({\bf v})$ is independent of the variable ${\bf x}$ present on the right side of Eq.~(\ref{X-defi}). We deal with a system which is assumed to be translationally invariant in the transverse plane but it is not fully uniform along the $z$-axis. We therefore expect a weak dependence of $X^{\alpha\beta} ({\bf v})$ on the longitudinal coordinate $z$ which is not explicitly shown on the left side of Eqs.~(\ref{Y-defi}) and~(\ref{X-defi}). 

The tensor $X^{\alpha\beta}({\bf v})$ in Eq.~(\ref{X-defi}) is expected to saturate at a large enough time $t$. This time independence occurs due to finite correlation lengths. If the correlator $\big\langle \mathcal{F}^\alpha(t, {\bf x}) \mathcal{F}^\beta (t', {\bf x}') \big\rangle$ vanishes for $|{\bf x}'- {\bf x}| > \lambda_x$ or $|t'- t| > \lambda_t$, the integral (\ref{X-defi}) saturates for $ t > \lambda_t$ or $ t > \lambda_x/v$. In practice, the saturation of $X^{\alpha\beta} ({\bf v})$ at long times and its approximate independence on $z$ provide an estimate of the range of validity of the approximations that we use to obtain the  transport coefficients of the glasma. We note that we have not indicated dependence on time on the left side of Eqs.~(\ref{Y-defi}) and (\ref{X-defi}) since the transport coefficients that we will calculate are only meaningful when at least approximate saturation is observed. We return to these points in Sec.~\ref{sec-results}.


The physical interpretation of the collision terms $Y^\alpha({\bf v})$ and $X^{\alpha\beta}({\bf v})$ is easy to understand. 
As discussed in the textbook \cite{Kampen:1987}, they determine the average momentum change per unit time, and the correlation of momentum changes per unit time, as follows
\ba
\label{Dpi-final}
\frac{\langle \Delta p^\alpha \rangle}{\Delta t} 
&=& - Y^\alpha({\bf v}) ,
\\[2mm]
\label{Dpi-Dpj-final}
\frac{\langle \Delta p^\alpha \Delta p^\beta \rangle}{\Delta t} 
&=& X^{\alpha\beta}({\bf v}) + X^{\beta\alpha}({\bf v}) .
\ea
The collisional energy loss $dE/dx$ and the transverse momentum broadening parameter $\hat{q}$ of a heavy quark in a glasma are obtained from the results in Eqs.~(\ref{Dpi-final}) and (\ref{Dpi-Dpj-final}) using the equations 
\ba
\label{enn}
\frac{dE}{dx} &=& \frac{v^\alpha}{v} \frac{\langle \Delta p^\alpha \rangle}{\Delta t} ,\\
\label{qq}
\hat q &=& \frac{1}{v} \Big( \delta^{\alpha\beta} -\frac{v^\alpha v^\beta}{v^2} \Big)
\frac{\langle \Delta p^\alpha \Delta p^\beta \rangle}{\Delta t} ,
\ea
where $v=|{\bf v}|$. Equations~(\ref{XY}), (\ref{enn}) and (\ref{qq}) give
\ba
\label{e-loss-X-Y}
\frac{dE}{dx} &=& - \frac{v}{T} \frac{v^\alpha v^\beta}{v^2} X^{\alpha\beta}({\bf v}), \\
\label{qhat-X-T}
\hat{q} &=&  \frac{2}{v} \Big(\delta^{\alpha\beta} - \frac{v^\alpha v^\beta}{v^2}\Big) X^{\alpha\beta}({\bf v}).
\ea
The quantity $X^{\alpha\beta}({\bf v})$ in Eq.~(\ref{X-defi}) is determined by correlators of components of the colour Lorentz force, and consequently by correlators of the glasma electric and magnetic fields as follows
\ba
\label{tensor-gen}
X^{\alpha\beta}({\bf v}) &=&
\frac{g^2}{2 N_c} \int_0^t dt' \Big[\big\langle E_a^\alpha(t, {\bf x}) E_a^\beta(t-t',{\bf y})\big\rangle 
+ \epsilon^{\beta \gamma \gamma'}v^\gamma 
\big\langle E_a^\alpha(t, {\bf x}) B_a^{\gamma'}(t-t',{\bf y})\big\rangle 
\\[2mm]  
&&~~~~
+ \epsilon^{\alpha \gamma \gamma'}v^\gamma 
\big\langle B_a^{\gamma'}(t, {\bf x}) E_a^\beta (t-t',{\bf y})\big\rangle
+ \epsilon^{\alpha \gamma \gamma'} \epsilon^{\beta \delta \delta'}v^\gamma v^\delta 
\big\langle B_a^{\gamma'}(t, {\bf x}) B_a^{\delta'}(t-t',{\bf y})\big\rangle
\Big], \nn
\ea
where ${\bf y} = {\bf x} - {\bf v}t'$. For future convenience we also define ${\bf v}=(v_\parallel,{\bf v}_\perp)$ and $v_\perp=|{\bf v}_\perp|$. In~ Sec.~\ref{CGC-all} we describe how to calculate the correlators of the chromodynamic fields, which provides  an analytic form for the tensor $X^{\alpha\beta}({\bf v})$. We comment that the correlators in (\ref{tensor-gen}) are not  gauge invariant. In principle, this problem could be remedied by inserting link operators between the two fields, as discussed in Ref.~\cite{Mrowczynski:2017kso}, but practically this procedure is difficult to realize. 
The gauge invariant implementation of our method is an interesting and open issue  
that is beyond the scope of this paper.

We  concentrate on two projections of the tensor $X^{\alpha\beta}({\bf v})$, which allow us to calculate the collisional energy loss and momentum broadening coefficient (see Eqs.~(\ref{e-loss-X-Y}) and (\ref{qhat-X-T})). However, as will be shown in Sec.~\ref{CGC-all}, we have calculated all of the electric and magnetic field correlators which means that we have an analytic result for the full tensor, and this expression could be used in different calculations. For example, one
could solve the Fokker-Planck equation and determine how the distribution functions of hard probes evolve in time.

\section{Correlators of electric and magnetic fields in glasma}
\label{CGC-all}

In this section we briefly describe the basic steps required to calculate the correlators of the electric and magnetic fields that enter the collision terms defined in Sec.~\ref{FP-eq}. Further details are available in our previous paper~\cite{Carrington:2020ssh}. 

Two heavy nuclei move along the $z$-axis  with the speed of light. 
Because of Lorentz contraction, the nuclei are infinitely thin in the longitudinal direction. They collide at $t=z=0$, and the gauge potential that describes the  
strongly interacting matter both before and after the collision  can be represented by the following ansatz~\cite{Kovner:1995ts,Kovner:1995ja} 
\bea
\label{ansatz}
&& A^+(x) = \Theta(x^+)\Theta(x^-) x^+ \alpha(\tau, {\bf x}_\perp),  \nn\\
&& A^-(x) = -\Theta(x^+)\Theta(x^-) x^- \alpha(\tau, {\bf x}_\perp), \\
&& A^i(x) = \Theta(x^+)\Theta(x^-) \alpha_\perp^i(\tau, {\bf x}_\perp)
+\Theta(-x^+)\Theta(x^-) \beta_1^i(x^-, {\bf x}_\perp)
+\Theta(x^+)\Theta(-x^-) \beta_2^i(x^+, {\bf x}_\perp)\, ,\nonumber 
\eea
which is written using both light cone and Milne coordinates\footnote{The light cone coordinates are defined by $x^\mu_{\rm lc}=(x^+,x^-,{\bf x}_\perp)$, where $x^\pm=(t\pm z)/\sqrt{2}$ and the transverse components are ${\bf x}_\perp=(x^1_\perp,x^2_\perp)$, which are labelled by $i,j \in (1,2)$. The Milne coordinates are ${\bf x}_\perp$, $\tau=\sqrt{2x^+x^-}$ and $\eta=\ln(x^+/x^-)/2$.}.
The theta functions separate the pre-collision and post-collision regions of spacetime. The potential $\beta^i_1(x^-,{\bf x}_\perp)$ is the pre-collision potential of the right moving nucleus and $\beta^i_2(x^+,{\bf x}_\perp)$ is the pre-collision potential of  the left moving nucleus. The glasma potentials produced after the collision are represented by $\alpha(\tau,{\bf x}_\perp)$ and $\alpha^i_\perp(\tau,{\bf x}_\perp)$ and they are smooth functions in the forward light-cone region. Because of boost invariance the glasma potentials do not depend on $\eta$.

\subsection{Pre-collision potentials}
\label{CGC}

The CGC effective theory relies on a separation of scales distinguishing regions of large and small nucleon momentum fraction, denoted x. 
The large-x partons, which are the valence quarks, act as colour sources for small-x gluons which are represented as classical fields. Because of time dilation, the valance quarks are effectively frozen and do not behave as dynamical fields. The right-moving nucleus is independent of the light cone time $x^+$ and is represented by an SU($N_c$) four-current $J^\mu(x^-,{\bf x}_\perp)=\delta^{\mu +}\rho(x^-,{\bf x}_\perp)$, where $\rho(x^-,{\bf x}_\perp)$ is the colour-charge density. The soft classical gluon four-potential is denoted by $\beta^\mu(x^-,{\bf x}_\perp)$, 
 and this potential is also independent of the light cone time $x^+$, or `static', because the four-current that produced it, is static. The gluon field satisfies the Yang-Mills (YM) equations 
\ba
\label{YM-eq}
[D_\mu,F^{\mu\nu}] = J^\nu,
\ea
where $[\dots,\dots]$ denotes a commutator, $D_\mu=\partial_\mu - ig \beta_\mu$ is the covariant derivative, and $F^{\mu\nu}=\frac{i}{g}[D^\mu,D^\nu]=\partial^\mu \beta^\nu-\partial^\nu \beta^\mu -ig [\beta^\mu,\beta^\nu]$ is the gluon field strength tensor with $g$ being the coupling constant. 
The amplitude of the field is big because of the large occupation numbers of the soft gluons.

To solve the YM equation we use the ansatz
\bea
\label{cov-ansatz}
\beta_{\rm cov}^\mu (x^-,{\bf x}_\perp) = \delta^{\mu +}\Lambda(x^-, {\bf x}_\perp)\, ,
\eea
which is conventionally called covariant gauge because it satisfies the equation $\partial_\mu \beta_{\rm cov}^\mu=0$. 
The only non-zero components of the field strength tensor are $ F_{\rm cov}^{i+}= \partial^i \Lambda $, and the YM equation reduces to
\bea
\label{LAMsoln}
-\nabla_{{\bf x}_\perp}^2 \Lambda(x^-,{\bf x}_\perp)  = \rho(x^-,{\bf x}_\perp).
\eea  
The solution can be written in the adjoint representation as
\bea
\label{use-gf}
\Lambda^a(x^-,{\bf x}_\perp) = \int d^2 {\bf y}_\perp \, G({\bf x}_\perp - {\bf y}_\perp) \rho^a(x^-,{\bf  y}_\perp),
\eea
where the Green's function satisfies the equation
$ \nabla_{{\bf x}_\perp}^2 G({\bf x}_\perp - {\bf y}_\perp) = -\delta^{(2)}({\bf x}_\perp - {\bf y}_\perp)$
and can be written in momentum and coordinate space as
\bea
\label{G-form}
\tilde G({\bf k}_\perp) = \frac{1}{{\bf k}_\perp^2+m^2}, \qquad\qquad
G({\bf x}_\perp) = \frac{1}{2\pi}K_0\big(m |{\bf x}_\perp| \big),
\eea
where $K_0$ is the modified Bessel function of the second type. In Eq. (\ref{G-form}), the parameter $m$ is an infra-red regulator, which is identified with $\Lambda_{\textrm{QCD}}$, so that $m \approx \Lambda_{\textrm{QCD}} \approx 200$ MeV. This choice naturally encodes the behaviour of confinement, since it ensures that the colour charges in a nucleon are neutralized at the length scale which coincides with $\Lambda_{\textrm{QCD}}^{-1}$. We note that the need for this regulator is a consequence of the fact that confinement does not emerge naturally from the CGC effective theory.

The pre-collision potentials are connected to the glasma potentials through boundary conditions, which are given at the end of this section. 
The form of the ansatz for the glasma potentials (\ref{ansatz}) dictates that these boundary conditions involve pre-collision potentials in light-cone gauge. We must therefore transform the pre-collision potentials we have just found in covariant gauge, into light-cone gauge. To find the appropriate gauge transformation, we  solve $\beta^+(x^-,{\bf x}_\perp)=0$ with
\bea
\beta^+(x^-,{\bf x}_\perp) = \frac{i}{g} U^\dagger(x^-,{\bf x}_\perp) \partial^+ U(x^-,{\bf x}_\perp) + U^\dagger(x^-,{\bf x}_\perp) \beta_{{\rm cov}}^+(x^-,{\bf x}_\perp) U(x^-,{\bf x}_\perp) \,.
\eea
The solution is
 \bea
 U(x^-,{\bf x}_\perp) 
 = {\cal P}{\rm exp}\Big[i g \int_{-\infty}^{x^-} dz^- \Lambda(z^-,{\bf x}_\perp)\Big]\, , \label{myU1}
 \eea
where the lower limit on the integral is chosen to give retarded boundary conditions and the notation $\cal{P} $ indicates path ordering with the `left-later' convention. 
The transverse components in light-cone gauge therefore satisfy 
\bea
\beta^i(x^-,{\bf x}_\perp) = \frac{i}{g} U^\dagger(x^-,{\bf x}_\perp) \partial^i U(x^-,{\bf x}_\perp) \,.
\label{pg1}
\eea

Equations~(\ref{use-gf}), (\ref{G-form}), (\ref{myU1}) and (\ref{pg1}) determine the light cone gauge pre-collision potential $\beta^i(x^-,{\bf x}_\perp)$ in terms of the source function $\rho(x^-,{\bf x}_\perp)$ of the right moving ion. The potential for the left moving ion is obtained in the same way using an ansatz that is independent of $x^-$. The two potentials will henceforth be written $\beta^i_1(x^-,{\bf x}_\perp)$ and $\beta^i_2(x^+,{\bf x}_\perp)$.

The pre-collision solutions provide the initial glasma fields $\alpha(0, {\bf x}_\perp)$ and $\alpha_\perp^i(0, {\bf x})$ through the boundary conditions
\begin{eqnarray}
\label{cond1}
\alpha^{i}_\perp (0,{\bf x}_\perp) &\equiv & \alpha^{i(0)}_\perp ({\bf x}_\perp) = \lim_{{\rm w}\to 0} \Big(\beta^i_1 (x^-,{\bf x}_\perp)+\beta^i_2 (x^+,{\bf x}_\perp) \Big),\\
\label{cond2}
\alpha (0,{\bf x}_\perp) &\equiv& \alpha^{(0)}({\bf x}_\perp)  = -\frac{ig}{2} \lim_{{\rm w}\to 0} \Big[\beta^i_1 (x^-,{\bf x}_\perp),\beta^i_2 (x^+,{\bf x}_\perp)\Big],
\end{eqnarray}
where the letter ${\rm w}$ denotes the longitudinal extent of each nucleus across the light-cone. This finite width is introduced for technical reasons and is taken to zero at the end of the calculation, so that the pre-collision potentials depend only on the transverse coordinates. These boundary conditions can be found by matching the YM equations in the pre-collision and post-collision regions \cite{Carrington:2020ssh}, and were originally obtained for infinitely thin nuclei in Refs.~\cite{Kovner:1995ts,Kovner:1995ja}.

\subsection{Glasma fields in the proper time expansion}
\label{sec-prop-time}

The YM equations that describe the gluonic matter produced in a heavy ion collision were solved numerically for the first time in a series of papers~\cite{Krasnitz:2001qu,Krasnitz:1999wc,Krasnitz:2000gz,Krasnitz:1998ns} and later on by several groups. Our calculations are analytic and therefore necessarily involve some simplifying assumptions that are not needed in a numerical approach. However, analytic methods always provide a valuable complement to numerical calculations. One advantage is that they allow for better control over various approximations and sources of errors than typical numerical simulations. 

The glasma potentials at finite proper time can be obtained from the initial glasma potentials using a proper time expansion, which is also called a ``near field expansion''. This method is based on the idea that since the lifetime of the glasma phase is very short, $\tau\lesssim 1$~fm, one can treat the proper time $\tau$ as a small parameter to power-expand the glasma fields \cite{Fries:2006pv,Chen:2015wia,Fries:2017ina}. The first serious attempt to study the radius of convergence of the expansion, in the specific case of a calculation of the energy-momentum tensor, can be found in Refs.~\cite{Carrington:2020ssh,Carrington:2021qvi}. Within this approach the glasma potentials are expanded as
\begin{eqnarray}
\alpha^i_\perp(\tau, {\bf x}_\perp) =
\sum_{n=0}^\infty \tau^n \alpha^{i(n)}_{\perp} ({\bf x}_\perp), \qquad\qquad
\alpha(\tau, {\bf x}_\perp) = \sum_{n=0}^\infty \tau^n \alpha^{(n)} ({\bf x}_\perp). \label{alpha-exp}
\end{eqnarray}
The YM equations for the expanded glasma potentials can be solved recursively so that higher order expansion coefficients are given in terms of the zeroth order coefficients, which are written in terms of the pre-collision potentials in Sec.~\ref{CGC}, and their derivatives. 

Using similar notation for the coefficients of the expanded electric and magnetic fields we write
\begin{eqnarray}
\label{E-tau}
{\bf E}(\tau,{\bf x}_\perp) &=& {\bf E}^{(0)}({\bf x}_\perp) + \tau {\bf E}^{(1)}({\bf x}_\perp) + \tau^2 {\bf E}^{(2)}({\bf x}_\perp) + \mathcal{O}(\tau^3),\\
\label{B-tau}
{\bf B} (\tau,{\bf x}_\perp)&=& {\bf B}^{(0)}({\bf x}_\perp) + \tau {\bf B}^{(1)}({\bf x}_\perp) + \tau^2 {\bf B}^{(2)}({\bf x}_\perp) + \mathcal{O}(\tau^3). 
\end{eqnarray}
At zeroth order the transverse components of both fields are zero and 
\ba 
\label{E0}
E^{z(0)}({\bf x}_\perp) &=&   -2 \alpha^{(0)}({\bf x}_\perp) , \\ 
\label{B0}
B^{z(0)}({\bf x}_\perp) &=&
\partial^y \alpha^{x(0)}_\perp({\bf x}_\perp) - \partial^x \alpha^{y(0)}_\perp({\bf x}_\perp)
-ig [\alpha^{y(0)}_\perp({\bf x}_\perp),\alpha^{x(0)}_\perp({\bf x}_\perp)].
\ea
The explicit forms of the fields in terms of the initial glasma potentials, up to fourth order in $\tau$, can be found in Appendix B of Ref.~\cite{Carrington:2020ssh}.

\subsection{Correlators of initial gauge potentials}
\label{Before-LC-corr}

The colour structure of each ion enters through the area charge density of the source $\mu({\bf x}_\perp)$, which is defined through the equations
\bea
\label{big-in}
\langle \rho_{a} (x^-,{\bf x}_\perp) \rho_{b}(y^-,{\bf y}_\perp)\rangle 
&=& \delta_{ab}\lambda(x^-,{\bf x}_\perp)\delta(x^--y^-)\delta^{(2)}({\bf x}_\perp - {\bf y}_\perp), \\[2mm]
 \int dx^-\,\lambda(x^-,{\bf x}_\perp) &\equiv& \mu({\bf x}_\perp),
\eea
where $\lambda(x^-,{\bf x}_\perp)$ is a volume charge density. 
The angle brackets $\langle \dots \rangle$ mean averaging over an ensemble of sources, and
Eq.~(\ref{big-in}) is a consequence of the assumption that these sources can be treated as Gaussian distributed random variables.

In general we need to calculate correlators not of the sources, but of pre-collision potentials, of the form
\ba
\label{corr-gen}
\langle \beta^i_{a1} \beta^j_{b1} \dots \beta^k_{c2} \beta^l_{d2} \dots  \rangle.
\ea
We use the Glasma Graph approximation \cite{Lappi:2015vta} which means that Wick's theorem, which properly speaking should only be applied to a product of sources, is used on a product of pre-collision potentials. 
Correlators of an odd number of  potentials vanish.
Correlators of an even number of potentials can be expressed as products of pairs of pre-collision correlators. 
One also assumes that potentials from different nuclei are uncorrelated, so that we only need to calculate the 2-point correlator
\ba
\label{field-corr}
\delta_{ab} B_n^{ij}({\bf x}_\perp,{\bf y}_\perp) = \lim_{{\rm w}\to 0} \langle \beta^i_{na}(x^\mp,{\bf x}_\perp) \beta^j_{nb}(y^\mp,{\bf y}_\perp)\rangle,
\ea
where $n \in \{1,2\}$ enumerates the two (right and left moving) ions and the upper/lower sign on the light-cone variables corresponds always to the first/second ion. This two point correlator was derived for the first time in Ref.~\cite{Jalilian-Marian:1996mkd}. Its detailed derivation can be also found in Appendix~D of our previous paper~\cite{Carrington:2020ssh}. The result is
\bea
&& B_n^{ij}({\bf x}_\perp,{\bf y}_\perp) = \frac{2 }{g^2 N_c \tilde\Gamma_n({\bf x}_\perp,{\bf y}_\perp)}  \; \left[{\rm exp} \bigg(\frac{g^4 N_c}{2}\;\tilde\Gamma_n({\bf x}_\perp,{\bf y}_\perp) \bigg) -1\right]\;
\partial_x^i \partial_y^j \tilde\gamma_n({\bf x}_\perp,{\bf y}_\perp), \, \;\;\;\;
\label{B-res} 
\eea
where the functions $\tilde \gamma ({\bf x}_\perp - {\bf y}_\perp)$ and $\tilde \Gamma ({\bf x}_\perp - {\bf y}_\perp)$ are 
\bea
\tilde\Gamma_n({\bf x}_\perp,{\bf y}_\perp) = 2\tilde\gamma_n({\bf x}_\perp,{\bf y}_\perp) - \tilde\gamma_n({\bf x}_\perp,{\bf x}_\perp) - \tilde\gamma_n({\bf y}_\perp,{\bf y}_\perp)  \label{Gamma-tilde-def}
\eea
and
\bea
&& \tilde \gamma_n({\bf x}_\perp,{\bf y}_\perp) = \int d^2 {\bf z}_\perp \, \mu_n({\bf z}_\perp)\, G({\bf x}_\perp-{\bf z}_\perp)\,G({\bf y}_\perp-{\bf z}_\perp) \,.
\label{gamma-tilde-def}
\eea

Some physical properties of the ions can be specified through the colour charge densities $\mu_1({\bf z}_\perp)$ and $\mu_2({\bf z}_\perp)$. An impact parameter can be introduced by shifting the centres of the distributions $\mu_1({\bf z}_\perp)$ and $\mu_1({\bf z}_\perp)$ in opposite directions. 
In this work we assume that the charge distributions of both ions are equal to each other and constant $\mu \equiv \mu_1({\bf z}_\perp)=\mu_2({\bf z}_\perp)$, so that the system is translation invariant in the transverse plane. 
The area density $\mu$ is proportional to the square of the saturation momentum scale $Q_s$, see for example \cite{Fukushima:2007ki,Lappi:2007ku}, but the determination of the proportionality factor requires methods beyond CGC. We use
\be
\label{Qs-vs-mu}
\mu = g^{-4} Q_s^2 .
\ee
The correlator (\ref{B-res}) takes the form
\be
\label{Ai0-Aj0-C1-C2}
B^{ij}(r)
= \delta^{ij}  C_1 (r) - \hat{r}^i \hat{r}^j C_2 (r) ,
\ee
where $r\equiv|{\bf x}_\perp - {\bf y}_\perp|$ and $\hat{r}^i \equiv r^i/r$. The functions $C_1(r)$ and $C_2(r)$ are
\ba
\label{C1-def}
C_1(r) &\equiv&  \frac{m^2 K_0 (mr)}{g^2 N_c  \big( mr K_1(mr) - 1\big) }
\bigg\{ \exp\bigg[\frac{g^4 N_c \mu \big( mr K_1(mr) - 1\big)  }{4\pi m^2 }\bigg] -1 \bigg\} ,
\\[2mm]
\label{C2-def}
C_2(r) &\equiv&  \frac{m^3 r \, K_1(mr) }{g^2 N_c  \big( mr K_1(mr) - 1\big) }
\bigg\{ \exp\bigg[\frac{g^4 N_c \mu \big( mr K_1(mr) - 1\big)  }{4\pi m^2 }\bigg] -1 \bigg\}.
\ea
The function $C_1(r)$ diverges logarithmically as $r \to 0$ and hence the correlator (\ref{Ai0-Aj0-C1-C2}) has to be regularized. This will be discussed in detail in Sec.~\ref{sec-results}, where we present our results.

\subsection{Field correlators}
\label{field-corr}

Using the techniques of the previous sections, we can now calculate all of the field correlators that enter the tensor $X^{\alpha\beta}({\bf v})$ in Eq.~(\ref{tensor-gen}). In Sec.~\ref{sec-prop-time} we described how to obtain the glasma electric and magnetic fields to arbitrary order in $\tau$ in terms of the initial glasma potentials $\alpha^{(0)}({\bf x}_\perp)$ and $\alpha^{i(0)}_\perp({\bf x}_\perp)$, and their derivatives. In Sec.~\ref{CGC} we gave the boundary conditions (\ref{cond1}) and  (\ref{cond2}) that connect the initial glasma potentials to the pre-collision potentials, and the solutions for the pre-collision potentials in terms of the ion sources (see Eqs.~(\ref{use-gf}), (\ref{G-form}), (\ref{myU1}) and (\ref{pg1})). In Sec~\ref{Before-LC-corr} we explained how to calculate correlation functions of pre-collision potentials. Combining all of these steps we can calculate the field correlators in terms of the functions $C_1(r)$ and $C_2(r)$ defined in Eqs.~(\ref{C1-def}) and (\ref{C2-def}). At lowest order, the correlators of the electric and magnetic fields are given by 
\ba
\label{EzEz-corr}
\langle E^{z(0)}_a({\bf x}_\perp) \, E^{z(0)}_b({\bf y}_\perp)\rangle  &=&  g^2 N_c \delta^{ab} 
\Big( 2 C_1^2 (r) - 2 C_1 (r) \, C_2 (r) + C_2^2 (r) \Big) ,
\\[2mm]
\label{BzBz-corr}
\langle B^{z(0)}_a({\bf x}_\perp) \, B^{z(0)}_b({\bf y}_\perp)\rangle  &=&  g^2 N_c \delta^{ab} 
\Big( 2 C_1^2 (r) - 2 C_1 (r) \, C_2 (r) \Big),
\\[2mm]
\label{EB}
\langle E^{z(0)}_a({\bf x}_\perp) \, B^{z(0)}_b({\bf y}_\perp)\rangle  &=& 0 .
\ea
All higher order correlators are given by similar expressions involving the functions $C_1(r)$ and $C_2(r)$ and their derivatives. 
The tensor $X^{\alpha\beta}({\bf v})$ also involves correlators of fields at the same point, and we treat these 
as 2-point correlators. 
We have obtained an analytic expression for the tensor $X^{\alpha\beta}({\bf v})$ up to order $\tau^5$ using {\it Mathematica}.

\section{Results - transport coefficients of hard probes in glasma}
\label{sec-results}

In this section we present our results.  
The collisional energy loss $dE/dx$ and the momentum broadening coefficient $\hat q$ of a hard  probe traversing a glasma are obtained from the tensor $X^{\alpha\beta}({\bf v})$ using equations (\ref{e-loss-X-Y}) and (\ref{qhat-X-T}). 
The tensor $X^{\alpha\beta}({\bf v})$, defined in Eq. (\ref{tensor-gen}), is given by correlators of  chromodynamic glasma fields which are calculated as described in Sec.~\ref{CGC-all}. 

\subsection{Introductory discussion}
\label{sec-res-0}

Heavy and high $p_T$ probes are produced at the very first moments of the collision and then propagate through the evolving glasma. When the magnitude of the velocity of the probe $v$ is close to one, it is highly relativistic, and could be a heavy or light quark, or gluon. When $v$ is less than one, the probe is necessarily a heavy quark. We will study only the behaviour of quark probes, but the tensor $X^{\alpha\beta}$ for a gluon, and consequently $\hat q$ and $dE/dx$, could be obtained from Eq.~(\ref{X-defi}) by multiplying by a factor $(N_c^2 - 1)/2N_c^2$, which equals $4/9$ for $N_c=3$. 

Because experiments at RHIC and the LHC focus on hard probes from the momentum space mid-rapidity region, $y\in (-1,+1)$, we are primarily interested in the transport properties of probes moving mostly perpendicularly to the beam axis. The momentum-space rapidity $y$ is related to the longitudinal component of the probe's velocity by $y=\frac{1}{2} \ln \frac{1+v_\parallel}{1-v_\parallel}$, and therefore the values $y=\pm 1$ correspond to $v_\parallel = \pm 0.76$, and the mid-rapidity value $y=0$ corresponds to strict transverse motion, $v_\parallel=0$. In this work we use the parameter $v_\parallel$ instead of $y$ to quantify deflection from transverse motion, and we consider quarks with $0\leq v_\parallel < 0.76$.
 
\begin{figure}[h]   
\centering
\includegraphics[scale=0.4]{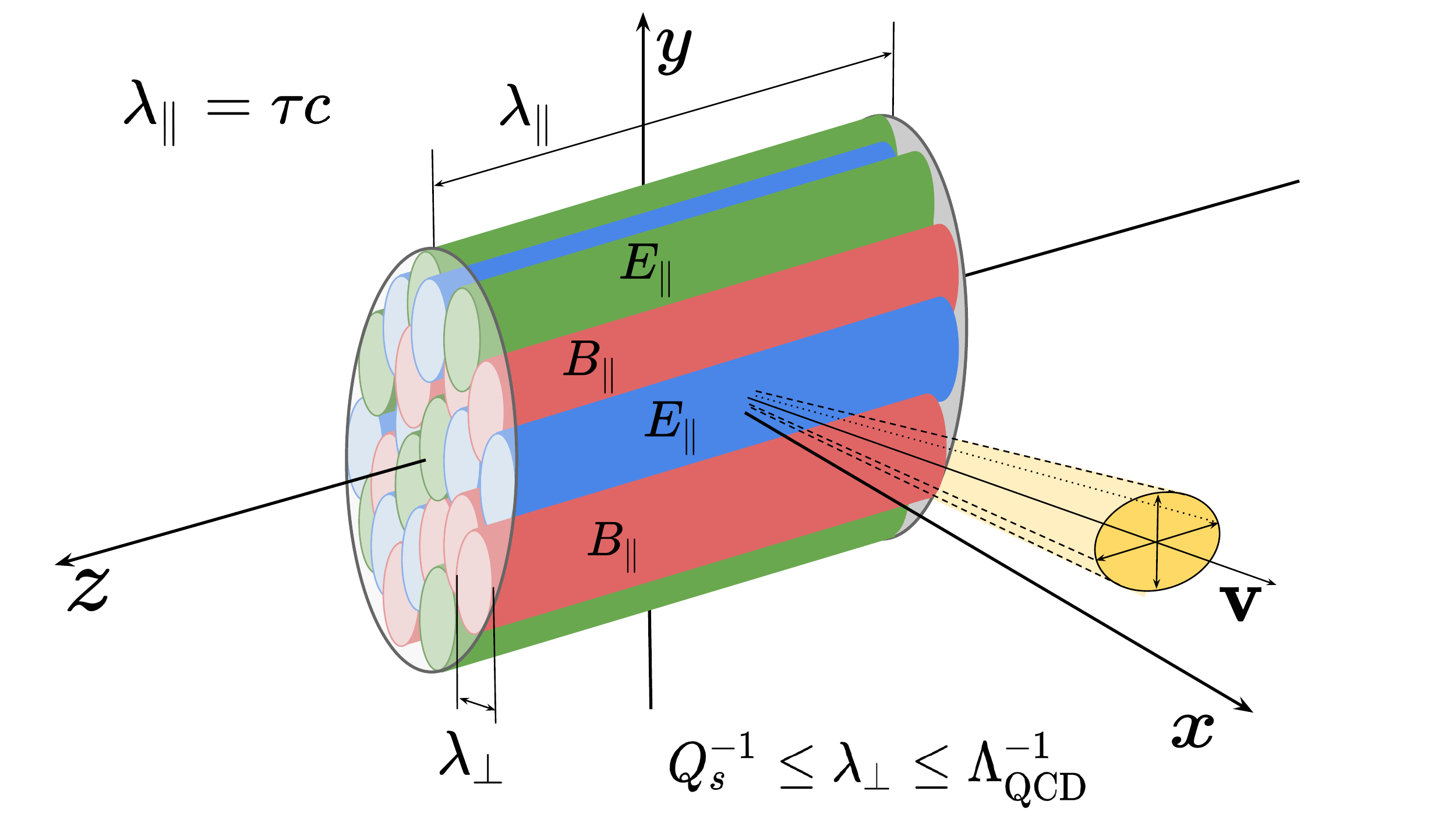}  
\caption{Cartoon of the zeroth order glasma fields and a probe moving mostly transverse to the collision axis.
\label{tubes}}  
\end{figure}

 
An idealized picture of a probe emerging from the glasma at very early proper times is shown in Fig.~\ref{tubes}, 
where the glasma fields at zeroth order in the proper time expansion are represented by coloured flux tubes. At this order the electric and magnetic fields are purely longitudinal and static. There are two qualitatively different correlation lengths, which we will denote $\lambda_\parallel$ and $\lambda_\perp$. The longitudinal correlation length $\lambda_\parallel$ is proportional to the distance between the nuclei and can be identified with the proper time $\tau$. The transverse correlation length $\lambda_\perp$ can be inferred from the correlators (\ref{EzEz-corr}) and (\ref{BzBz-corr}). Qualitatively the transverse correlation length obeys $Q_s^{-1} \leq\lambda_\perp \leq \Lambda^{-1}_{\rm QCD}$. 


The collisional energy loss and the momentum broadening parameter are both built up during the time that the probe spends within the domain of correlated fields. At zeroth order, this time is determined by the transverse correlation length and the orientation and magnitude of the probe’s velocity. The transport coefficients will saturate when the probe leaves the region of correlated fields\footnote{We note that we consider only the glasma phase, where coherent fields are present and correlation lengths are sizable. We do not consider later kinetic or hydrodynamic stages where further broadening of the momenta of hard probes occurs due to scattering on plasma constituents.}. 
These simple arguments indicate that the Fokker-Planck methodology we are using might be well suited to describe the problem of a hard probe moving through a glasma, at least at very early times.

The simple picture presented in Fig. \ref{tubes} is valid at zeroth order, but at later times it does not accurately describe the glasma. 
As $\tau$ increases, transverse electric and magnetic fields develop, and the glasma fields can grow or decrease rapidly. 
Higher and higher orders in the $\tau$ expansion are needed to describe the glasma fields, as $\tau$ increases.
Our method will work  if saturation is reached before the $\tau$ expansion that is used to calculate the field correlators breaks down.




In addition to determining how long a probe spends in the region of correlated fields, the probe's velocity affects the transport coefficients in another way. To see this we look at Eqs.~(\ref{X-defi}), (\ref{e-loss-X-Y}) and (\ref{qhat-X-T}), and use the form of the Lorentz force. At zeroth order the integrand that gives collisional energy loss is proportional to
\ba 
\label{struc-a}
v^\alpha v^\beta {\cal F}^{\alpha(0)} {\cal F}^{\beta(0)}&=& 
g^2 v^2_\parallel E^{z(0)}({\bf x}_\perp) E^{z(0)}({\bf x}'_\perp) ,
\ea
and the integrand for momentum broadening is proportional to 
\ba
\label{struc-b}
\Big(\delta^{\alpha\beta}-\frac{v^\alpha v^\beta}{v^2}\Big){\cal F}^{\alpha(0)} {\cal F}^{\beta(0)}
&=& 
g^2\frac{v_\perp^2}{v^2} \Big( E^{z(0)}({\bf x}_\perp) E^{z(0)}({\bf x}'_\perp)
+ v^2 B^{z(0)}({\bf x}_\perp) B^{z(0)} ({\bf x}'_\perp)\Big).
\ea
We see that at zeroth order the collisional energy loss is caused by the electric field and vanishes when the probe moves in the transverse direction. In contrast, zeroth order momentum broadening is caused by both electric and magnetic fields, and is maximal when the probe moves transversely. We will show that at higher orders the same behaviour is observed: for fixed $v$, when $v_\perp$ increases and $v_\parallel$ decreases, one finds that the collisional energy loss decreases and the momentum broadening increases.

We emphasize that the arguments presented in this section give a good qualitative description of the behaviour of the glasma at very early times, corresponding to the lowest orders of the $\tau$ expansion. Collisional energy loss and momentum broadening at
higher orders in the $\tau$ expansion require calculations to understand the full picture.

In Sec.~\ref{CGC-all} we explained that the correlators of chromodynamic fields that enter the tensor $X^{\alpha\beta}({\bf v})$ are restricted to the forward light-cone region, where the glasma description is valid. To take this condition into account in our calculation of the transport coefficients, the integrands of the tensor $X^{\alpha\beta}({\bf v})$ in Eq.~(\ref{tensor-gen}) should be multiplied by
\ba
\label{step}
\Theta(t^2-z^2) \Theta((t-t')^2-(z-v_\parallel t')^2).
\ea 
If we use $v_\parallel=0$ and look at $z=0$, both step functions can be ignored as they are always unity. When $v_{\parallel}$ is nonzero the second step function in (\ref{step}) has the effect of reducing $\hat q$ and $dE/dx$. 
In almost all calculations, we will choose $z=0$ so that the first step function plays no role. 
The exception is section \ref{sec-res-b} where we study the dependence of our results on spatial rapidity.

As discussed in Sec.~\ref{FP-eq}, in order to calculate $dE/dx$ we need the temperature $T$ of an equilibrated quark-gluon plasma, whose energy density is the same as the energy density of the glasma. The energy density of an equilibrium free quark-gluon plasma of $N_f$ flavours equals
\be
\label{en-den-QGP}
\varepsilon_{\rm QGP} = \frac{\pi^2}{60} \big( 4(N_c^2 -1) + 7 N_f N_c \big)  T^4 ,
\ee
where only quarks with masses much smaller than the temperature are included.
The effective temperature of the glasma can therefore be estimated from the glasma energy density which was calculated in our previous paper \cite{Carrington:2020ssh}. We have shown that to sixth order in the proper time expansion the energy density has the form
\be
\label{epsilon-glasma}
\varepsilon_{\rm QGP} = 130.17 \big(15.9773 - 29.6759 \, {\tilde{\tau}}^2 
+ 42.6822 \, {\tilde{\tau}}^4 - 49.2686 \, {\tilde{\tau}}^6 \big) ,
\ee
where $\tilde{\tau} \equiv Q_s \tau$ and the energy density is expressed in ${\rm GeV/fm^3}$. 
The magnitude of the coefficients in this result increase, which indicates that the expansion will break down as $\tau$ increases. The fact that the signs of the coefficients alternates is typical in perturbative calculations, and delays the breakdown of the expansion. 
The sixth order result in equation (\ref{epsilon-glasma}) is reliable to approximately $\tau=0.05$~fm (see figure 1 in Ref.  \cite{Carrington:2020ssh}). 
We will argue below that our calculation of the momentum broadening coefficient $\hat q$ is valid to $\tau\sim0.07$~fm, which appears to be inconsistent, but since the temperature depends on the 1/4 power of the energy density it is very insensitive to higher order contributions from the proper time expansion.

All our results are calculated for $N_c=3$ and $g=1$. In Secs.~\ref{sec-res-a}, \ref{sec-res-b} and \ref{sec-res-c} we use $Q_s=2$ GeV and $m=0.2$ GeV. In Sec.~\ref{sec-res-d} we consider different values of $Q_s$ and $m$. In Sec.~\ref{sec-res-b} we study the $\eta$ dependence of $\hat q$. In all other calculations we work at mid-spatial-rapidity, or $\eta=z=0$.

\subsection{Time dependence of $\hat q$ and $dE/dx$}
\label{sec-res-a}

\begin{figure}[h]   
\centering
\includegraphics[scale=0.195]{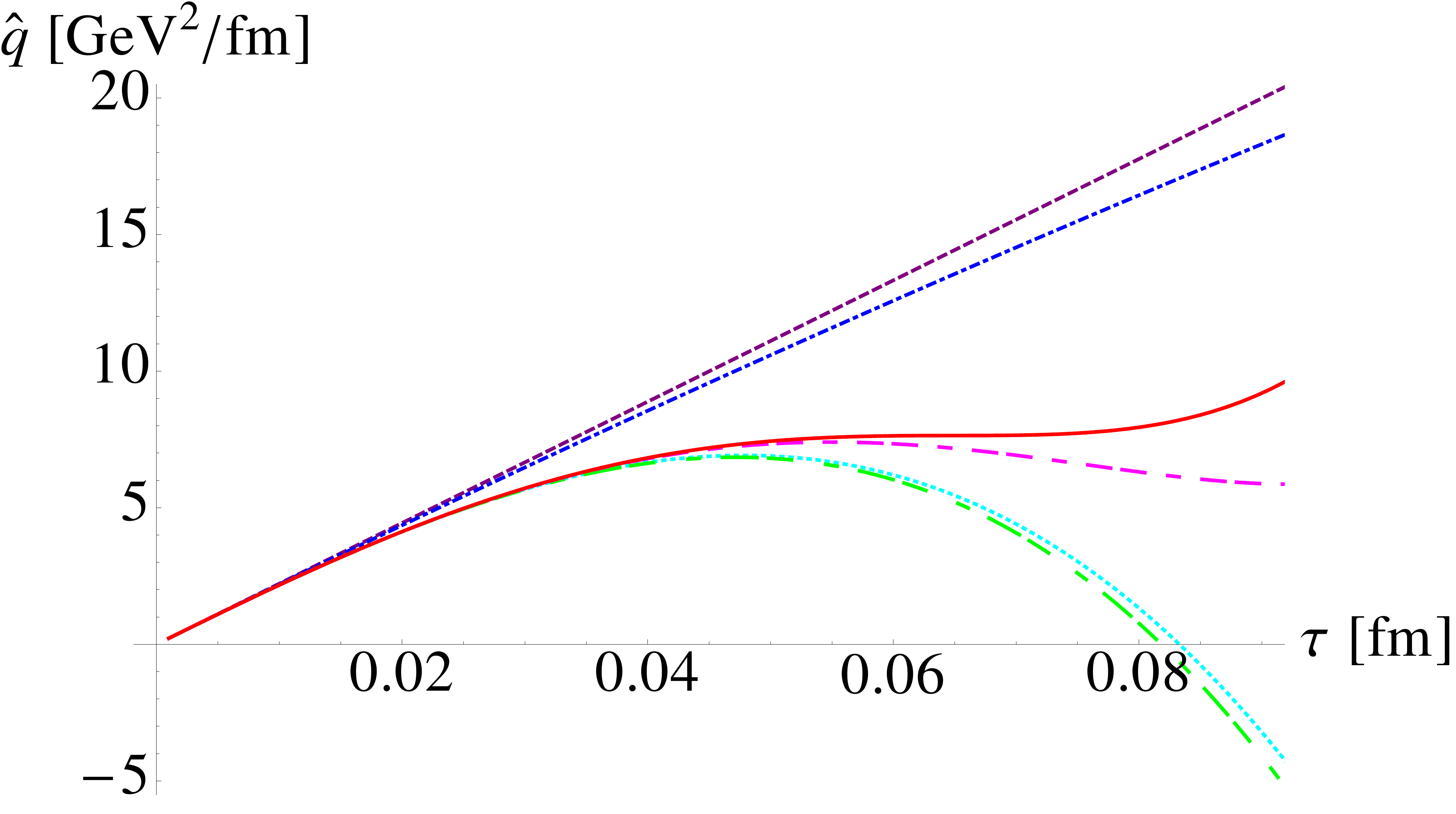} 
\includegraphics[scale=0.19]{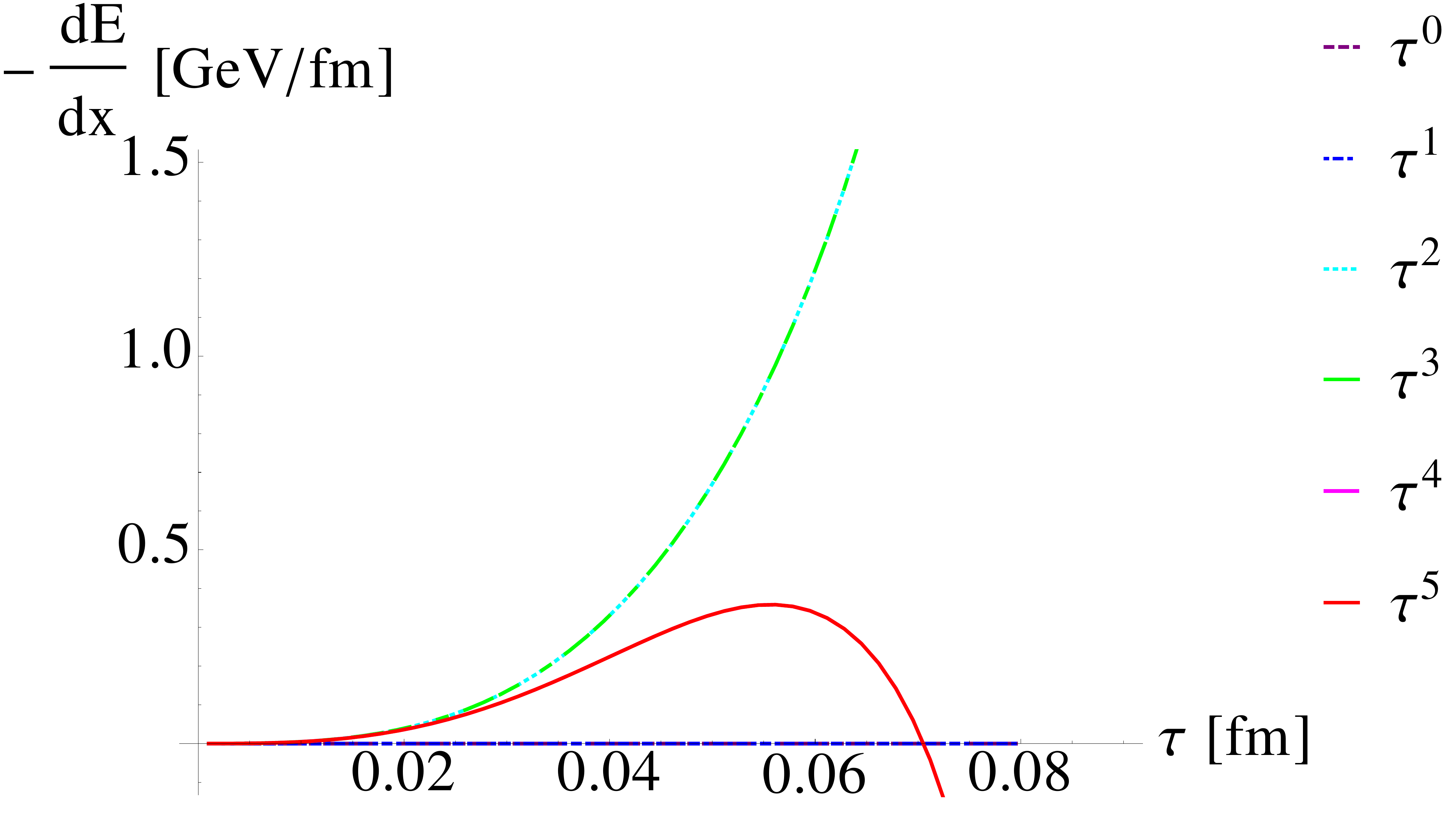}    
\caption{Transport coefficients of a ultra-relativistic quark, with $v=v_\perp=1$. Left panel: time evolution of the momentum broadening coefficient. Right panel: time evolution of the collisional energy loss. The results are shown at cumulative orders of $\tau$, see text for discussion.  
\label{t-coeff-cum}} 
\end{figure}

The momentum broadening coefficient $\hat q$ and the collisional energy loss $dE/dx$, both for an  ultra-relativistic hard probe moving perpendicularly to the beam axis, with $v=v_\perp=1$, are presented as a function of $\tau$ in Fig.~\ref{t-coeff-cum}. The left panel shows the momentum broadening coefficient and the right panel shows the collisional energy loss. The dependence of both transport coefficients on the order of the $\tau$ expansion is presented to illustrate the convergence of the expansion. For example, the dashed (purple) line in both panels represents the zeroth order contribution and the solid (red) line shows the $\tau^5$ cumulative results (all terms up to order $\tau^5$ order are summed). 

Taking into account higher order contributions in the $\tau$ expansion extends the range of validity of the result, 
which can be estimated from the largest value of $\tau$ for which the result at a given order agrees fairly well with the result at the previous order. 
At very early times, all orders of the $\tau$ expansion agree well. 
When all terms up to order $\tau^5$ are included, the time evolution of $\hat q$ shows initial growth, and then flattening, followed by more rapid growth (which appears after $\tau \sim 0.09$ fm). 
The region where $\hat q$ flattens shows saturation. 
The rapid increase of $\hat q$ at later times is not physical, but reflects the breakdown of the proper time expansion. 
At order $\tau^5$, the highest value of $\hat q$ that is obtained before the proper time expansion breaks down is about 6 GeV$^2$/fm. 
The coefficient $\hat q$ was also calculated in Ref.~\cite{Ipp:2020nfu} using real time QCD simulations. The time evolution of $\hat q$ found in this work is qualitatively similar to our finding. Our result is smaller, but still of comparable size. 

\begin{figure}[h]  
\centering
\includegraphics[scale=0.25]{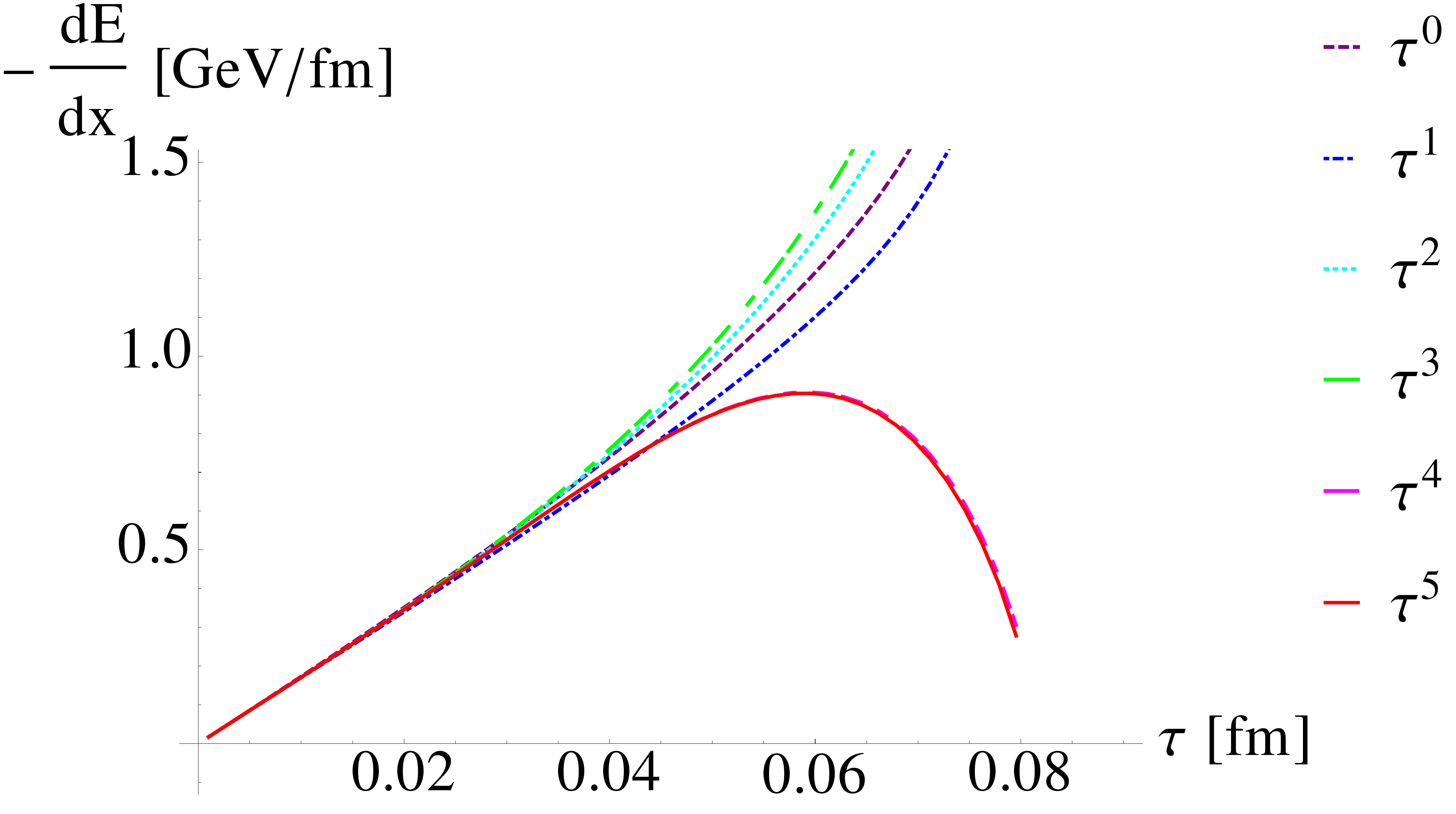}  
\caption{Time evolution of $dE/dx$ for $v=1$ and $v_\parallel=v_\perp=1/\sqrt{2}$.
\label{eloss-vel}}
\end{figure}

The behaviour of the collisional energy loss $dE/dx$ (depicted in the right panel of Fig.~\ref{t-coeff-cum}) is very different from what is seen from $\hat q$. 
Only the terms at $\tau^2$ and $\tau^4$ order contribute to the final result. All other orders vanish because they are proportional to some power of $v_\parallel$, which is zero in the case shown in Fig.~\ref{t-coeff-cum}. The collisional energy loss increases up to around $\tau=0.05$ fm, where it reaches a maximal value of approximately $0.4$~GeV/fm, and for larger times the expansion rapidly breaks down. 
From the discussion in Sec.~\ref{sec-res-0} we know that the collisional energy loss is much more sensitive to the value of longitudinal component of the velocity than $\hat q$ is, because the leading order contributions are proportional to $v_\parallel$. In this sense the case of purely transverse motion in Fig.~\ref{t-coeff-cum} might not represent typical behaviour. We therefore show $dE/dx$ in Fig.~\ref{eloss-vel}, also for $v=1$, but now with $v_\perp=v_\parallel=1/\sqrt{2}$. The shape of $dE/dx$ is not significantly different from the one shown in Fig.~\ref{t-coeff-cum}, but all 
orders in the $\tau$ expansion contribute to the final result.
The collisional energy loss is noticeably bigger and equals approximately 0.9 GeV/fm at its maximum, at around $\tau = 0.05$ fm. The $\tau$ expansion breaks down soon after this point. The absence of clear evidence of saturation indicates that our results for collisional energy loss should be considered  order of magnitude estimates only.

We comment that even for the momentum broadening coefficient, the effect of saturation should be studied carefully. In Ref. \cite{Carrington:2020sww} we calculated $\hat q$ and $dE/dx$ at order $\tau^1$, and both transport coefficients 
appeared to saturate in a striking manner. However, the higher order calculations presented in this paper show that the effect was seen at times that are far beyond the region of validity of the proper time expansion. When higher orders in the $\tau$ expansion are included, we see that the saturation observed at the first two orders in Ref. \cite{Carrington:2020sww} was an artifact of the approximation.

\subsection{Dependence of $\hat q$ on velocity and space-time rapidity}
\label{sec-res-b}


In the first part of this section we explore the dependence of $\hat q$ on the probe's velocity. 
We want to understand the relative importance of the two velocity dependent effects discussed in Sec.~\ref{sec-res-0}: the amount of time the probe spends in the region of correlated fields, and the dependence of the Lorentz force on the direction of the probe's velocity.  Next, we study the $\eta$ dependence of $\hat q$.

\begin{figure}[!h]
\centering
\includegraphics[scale=0.25]{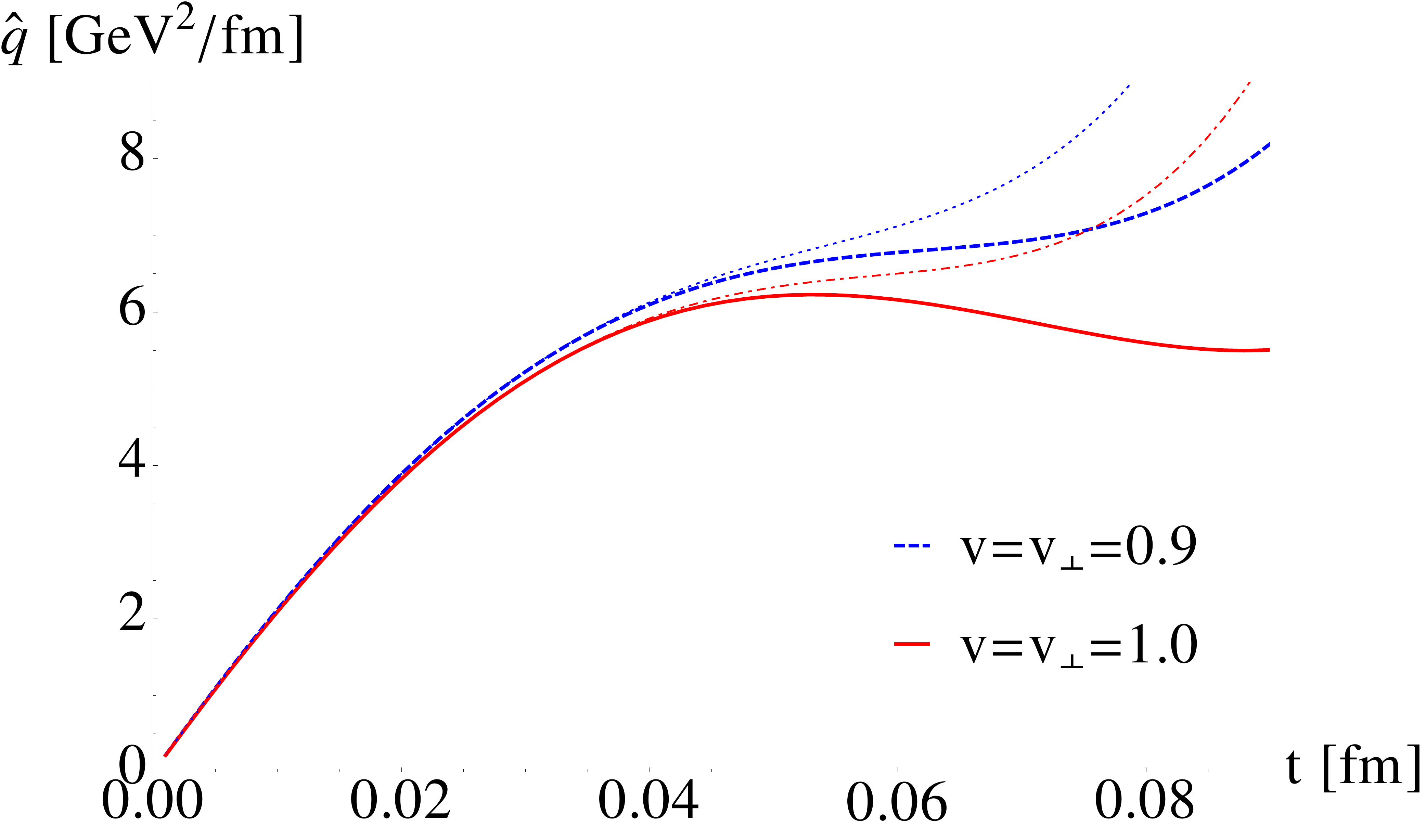}   
\caption{The time evolution of $\hat q$ at order $\tau^5$ for different values of $v=v_\perp$ is presented. Two cases are studied: $v=v_\perp=1$ denoted by the solid red line and $v=v_\perp=0.9$ denoted by the dashed blue line. The dotted and dashed-dotted lines represent the  corresponding results at order $\tau^4$.
\label{velocity}}    
\end{figure}

In Fig.~\ref{velocity} we show the dependence of $\hat q$ on the speed of a hard probe when the probe moves in the transverse direction. The solid (red) line represents the momentum broadening coefficient for a probe with $v=v_\perp=1$ and the dashed (blue) line is $v=v_\perp=0.9$. The thin dotted and dashed-dotted lines correspond to the results calculated at order $\tau^4$, which are shown to indicate the region of $\tau$ where the expansion converges. We observe that at very early times the momentum broadening coefficient is largely independent of $v=v_\perp$, but differences appear at longer times. The value of $\hat q$ for slower quarks flattens and then starts rapidly growing again, whereas $\hat q$ for ultra-relativistic quarks slightly decreases. This shows that when the transverse velocity of the probe increases at fixed $v_\parallel=0$, even though the Lorentz force contribution to $\hat q$ increases (see Eq.~(\ref{struc-b})), the dominant effect is the reduction of the amount of time the probe spends in the domain of correlated fields, which results in a reduction in momentum broadening.
This result agrees with the findings of Ref.~\cite{Ipp:2020nfu}, where the momentum broadening parameter of massless quarks is consistently smaller than for larger mass quarks, throughout the whole time evolution. 

\begin{figure}[h]
\centering
\includegraphics[scale=0.19]{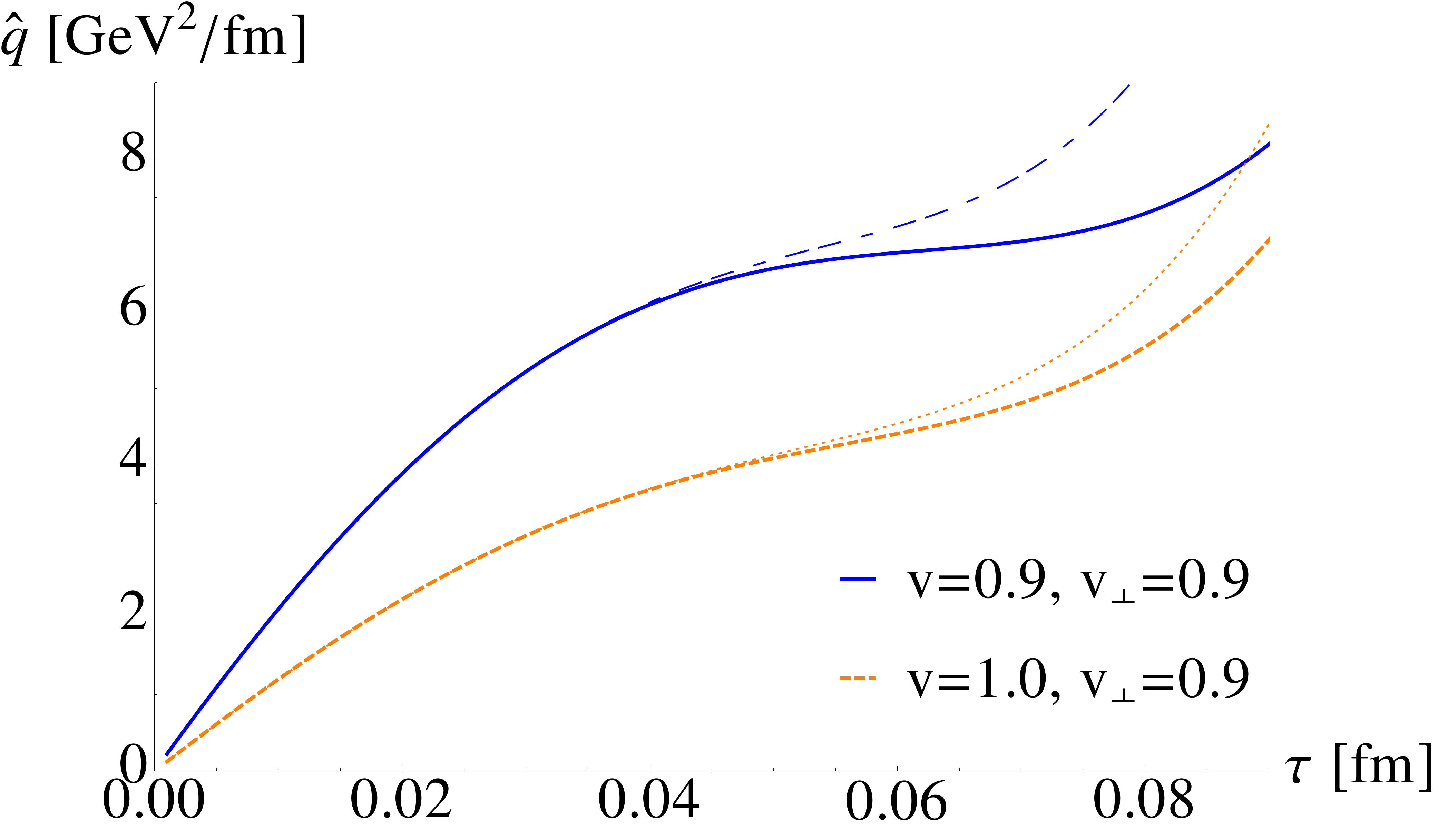}  
\includegraphics[scale=0.19]{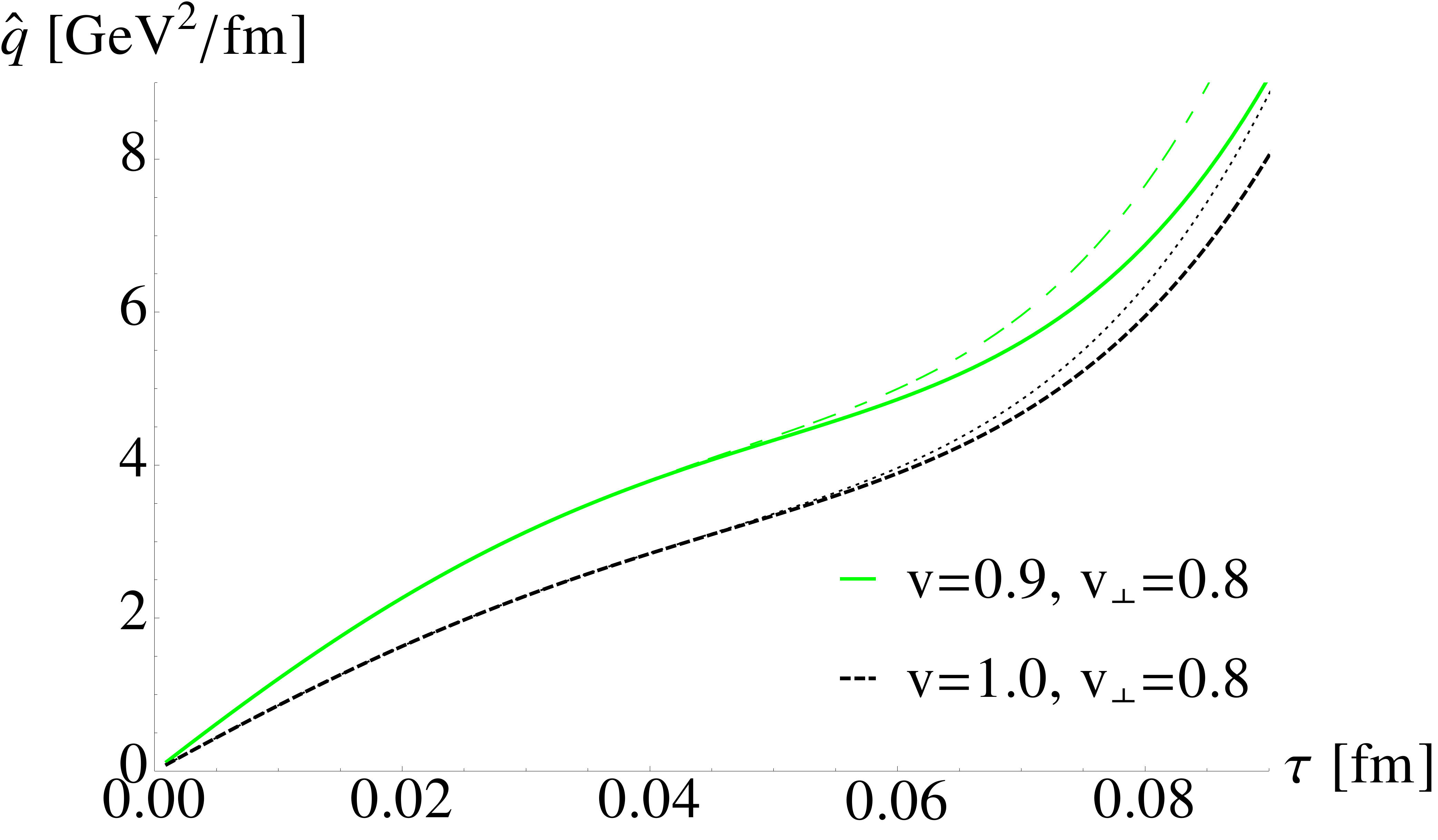}    
\caption{The time evolution of $\hat q$ with non-zero longitudinal component of the velocity. The left and right panels show $\hat q$ with $v_\perp=0.9$ and $v_\perp=0.8$, respectively.
\label{two-cases-par}}
\end{figure}

\begin{figure}[h]
\centering
\includegraphics[scale=0.19]{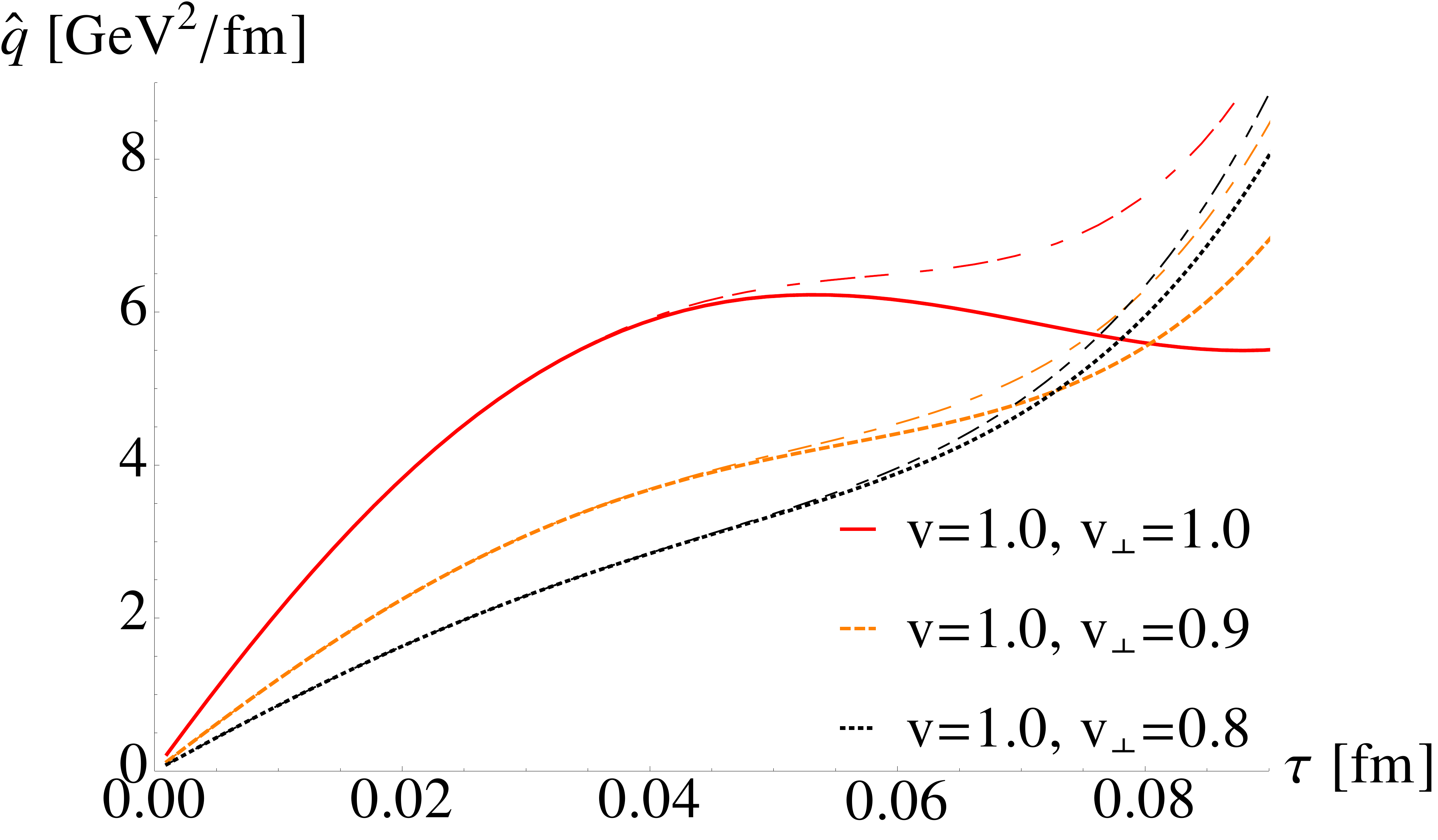}  
\includegraphics[scale=0.19]{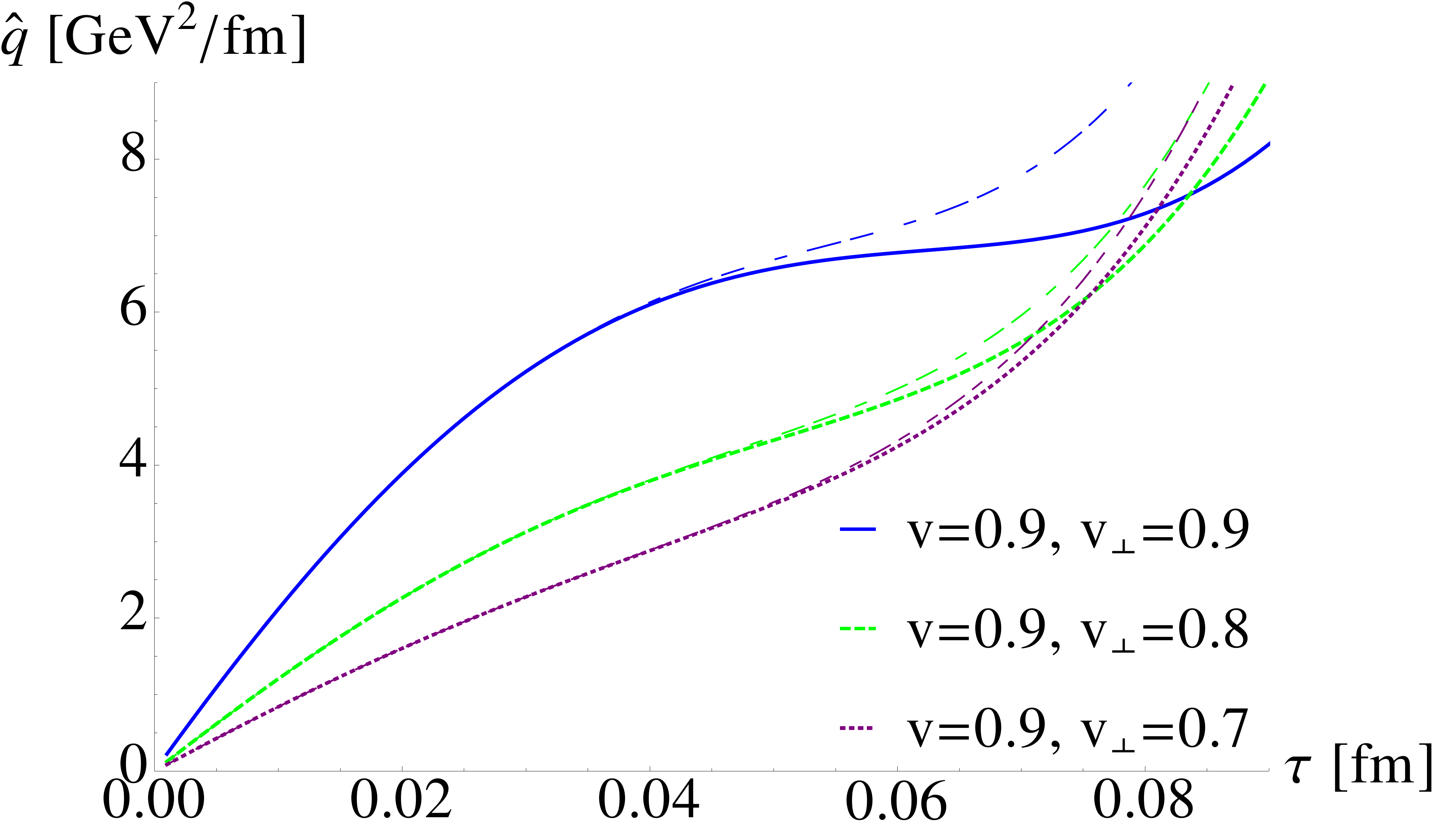}    
\caption{The time evolution of $\hat q$ with non-zero longitudinal component of the velocity. The left and right panels show $\hat q$ with $v=1$ and $v=0.9$, respectively.
\label{two-cases}}
\end{figure}

The radius of convergence of the $\tau$ expansion in our calculation can be estimated by comparing $\hat q$ at different orders in $\tau$ for each velocity. For both values of $v=v_\perp$ that we considered, the result for $\hat q$ can be trusted to $\tau$ around $0.07-0.08$ fm.
In Figs.~\ref{two-cases-par} and \ref{two-cases} we show the momentum broadening parameter for several cases with non-zero $v_\parallel$. In all cases we show the result at order $\tau^5$ (thick lines) and $\tau^4$ (thin lines). We consider two different values of the speed $v$. 
For each value of $v$ we consider the case of purely transverse motion, and several different non-zero values of the longitudinal velocity. 

The results for $\hat q$ with the same $v_\perp$ and different $v$ are shown in Fig.~\ref{two-cases-par}: in the left panel the results are calculated with $v_\perp=0.9$ and in the right panel with $v_\perp=0.8$. 
In the left panel of Fig.~\ref{two-cases} we show $\hat q$ for $v=1$ and transverse velocity components: $v_{\perp}=1$ (red line), $v_\perp=0.9$ (orange line) and $v_\perp=0.8$ (black line); these correspond to $v_\parallel=0$, $v_\parallel=0.44$ and $v_\parallel=0.6$, respectively. 
In the right panel, $\hat q$ is shown for $v=0.9$ and transverse components: $v_\perp=0.9$ (blue line), $v_\perp=0.8$ (green line) and $v_\perp=0.7$ (purple line), which correspond to: $v_\parallel=0$, $v_\parallel=0.41$ and $v_\parallel=0.57$, respectively.  (Lines with the same colour in Figs.~\ref{two-cases-par} and \ref{two-cases} denote the same values of $v$ and $v_\perp$.)

In all of the cases considered in Figs.~\ref{two-cases-par} and \ref{two-cases}, the results calculated at orders $\tau^4$ and $\tau^5$ agree quite well up to about $\tau \sim 0.07-0.08$ fm. 
The perpendicular component of the velocity $v_\perp$ is fixed in both panels of Fig.~\ref{two-cases-par}, and therefore the curves show $\hat q$ for two probes that spend the same amount of time in the region of correlated fields. The difference in the curves shows the effect of the velocity dependence of the Lorentz force. From Eq.~(\ref{struc-b}) one sees that the magnetic contribution is proportional to $v^2_\perp$ and the electric one to $v^2_\perp/v^2$, meaning that the role of electric contribution decreases when $v_\parallel$ increases at fixed $v_\perp$.
In Fig.~\ref{two-cases} we fix $v$ and vary $v_\perp$, which means we now also include the effect of changing the amount of time the probe spends in the region of correlation. In this case we see that at small times the probe with larger $v_\perp$ has larger $\hat q$, due to the larger Lorentz force, but 
as $\tau$ increases the curves with large and small $v_\perp$ cross each other. 
This happens because probes with larger $v_\perp$ escape from the region of correlated fields before the glasma fields become very large, but probes with smaller $v_\perp$ (and larger $v_\parallel$) remain in the domain of correlated fields for a longer time and eventually interact with very large fields. We also note that when $v_\perp=v$, the $\tau$ expansion breaks down at approximately the same point that the saturation regime disappears, but when $v_\perp \approx 0.9 v$, or smaller, the $\tau$ expansion converges fairly well even though no significant saturation regime is observed. This suggests that including higher orders in the $\tau$ expansion could extend the region of saturation when $v_\perp=v$.



\begin{figure}[h] 
\centering
\includegraphics[scale=0.19]{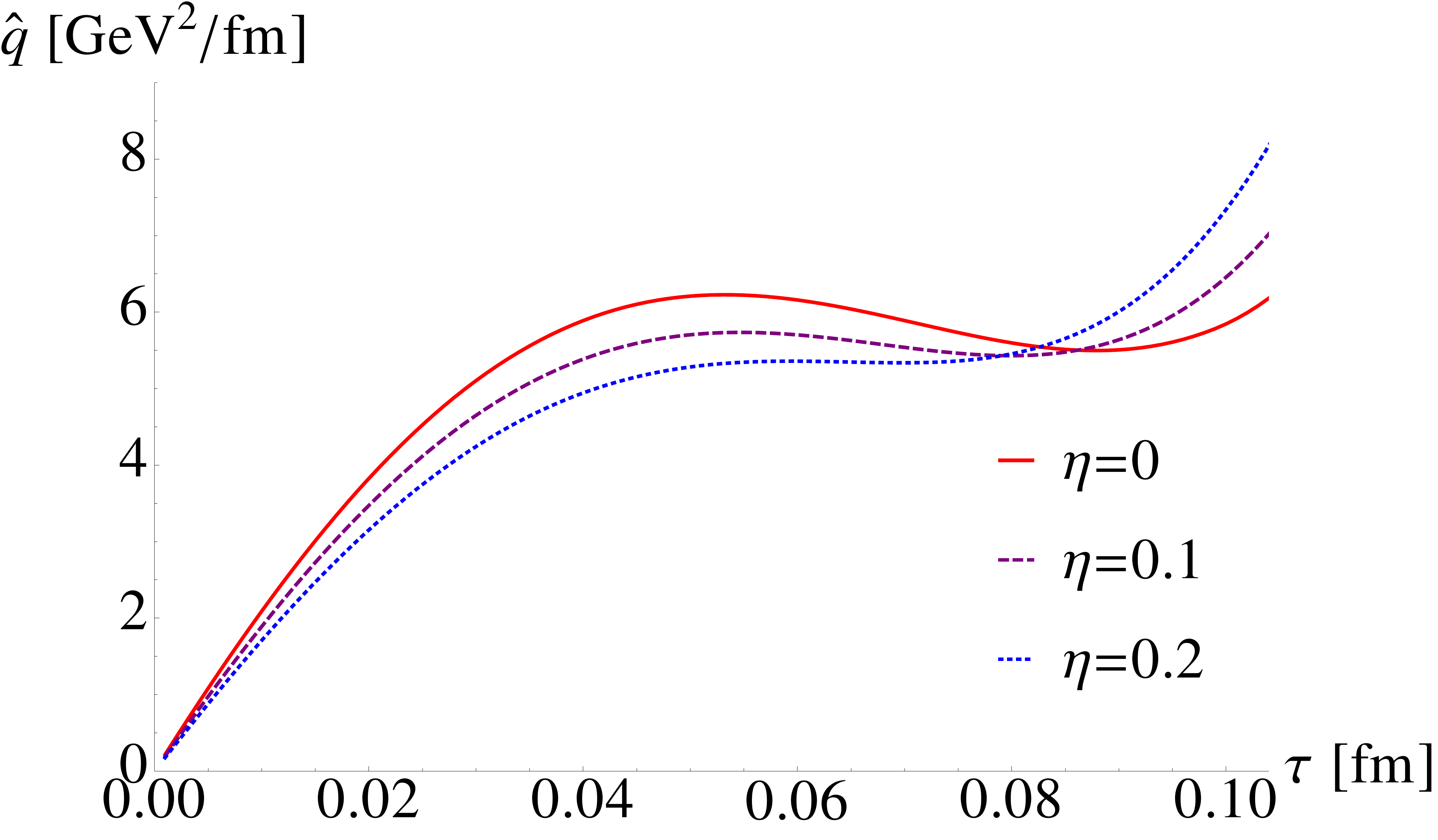}  
\includegraphics[scale=0.19]{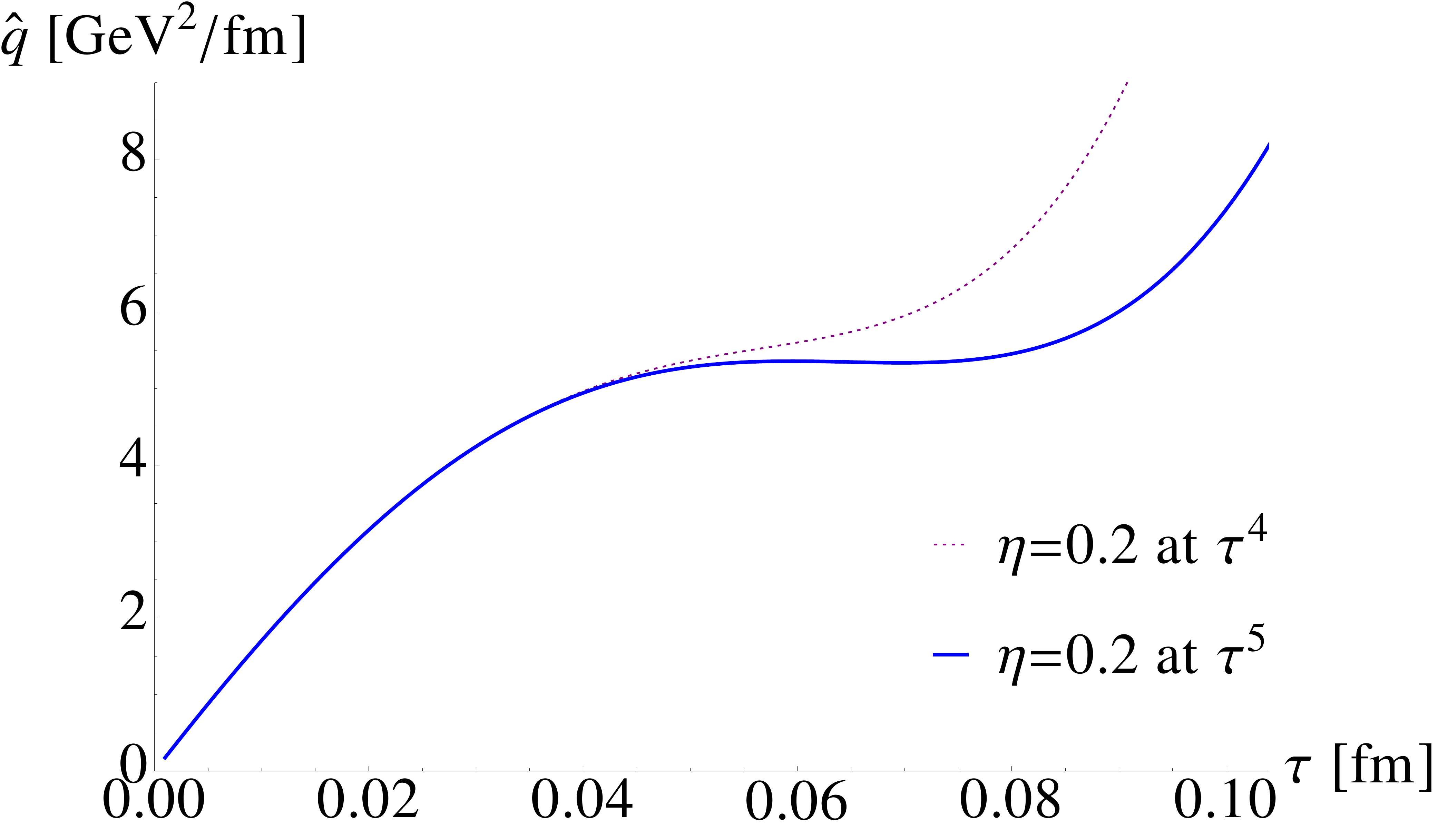}    
\caption{Dependence of $\hat q$ on the spatial rapidity $\eta$ calculated for the velocity $v=v_\perp=1$. 
\label{rapidity-plot}}
\end{figure}

In Fig.~\ref{rapidity-plot} we show the dependence of $\hat q$ on the spatial rapidity $\eta$, which is related to the initial position of the probe on the $z$ axis. 
In the left panel, $\hat q$ is displayed as a function of $\tau$ for three values of $\eta$. In the right panel the results for $\hat q$ with $\eta=0.2$ are depicted at order $\tau^4$ and order $\tau^5$  to indicate that this result can be trusted to $\tau \sim 0.06$ fm.

We consider only small values of $\eta$ because our approach is expected to work best in the mid-spatial-rapidity region, where the CGC approach that we use is most reliable. 
The momentum broadening parameter depends only weakly on $\eta$ in the region where the curves flatten. The figure shows that at least until the region of approximate saturation ends, the result for $\hat q$ is largely independent of of spatial rapidity. This result verifies that there is a range of proper times for which the boost invariant ansatz that was used to calculate the glasma correlators in Sec.~\ref{CGC-all} is compatible with the approximations that were used to derive the Fokker-Planck equation in Sec. \ref{FP-eq}. 

\subsection{Dependence on IR and UV energy scales}
\label{sec-res-d}

The UV scale $Q_s$ and the IR regulator $m$ enter our calculation as parameters that are related to the saturation and confinement scales. We remind the reader of the physical picture: we must stay below $Q_s$, or the assumption that the glasma is composed of classical gluon fields breaks down, and above $m$, so that we do not enter the regime where non-perturbative effects become dominant. The CGC approach is valid only within this relatively narrow window. The numerical values of these scales cannot be precisely determined within the formalism we are using. It is therefore important to see how varying these parameters influences the time evolution of $\hat q$. 

\begin{figure}[h]
\centering 
\includegraphics[scale=0.19]{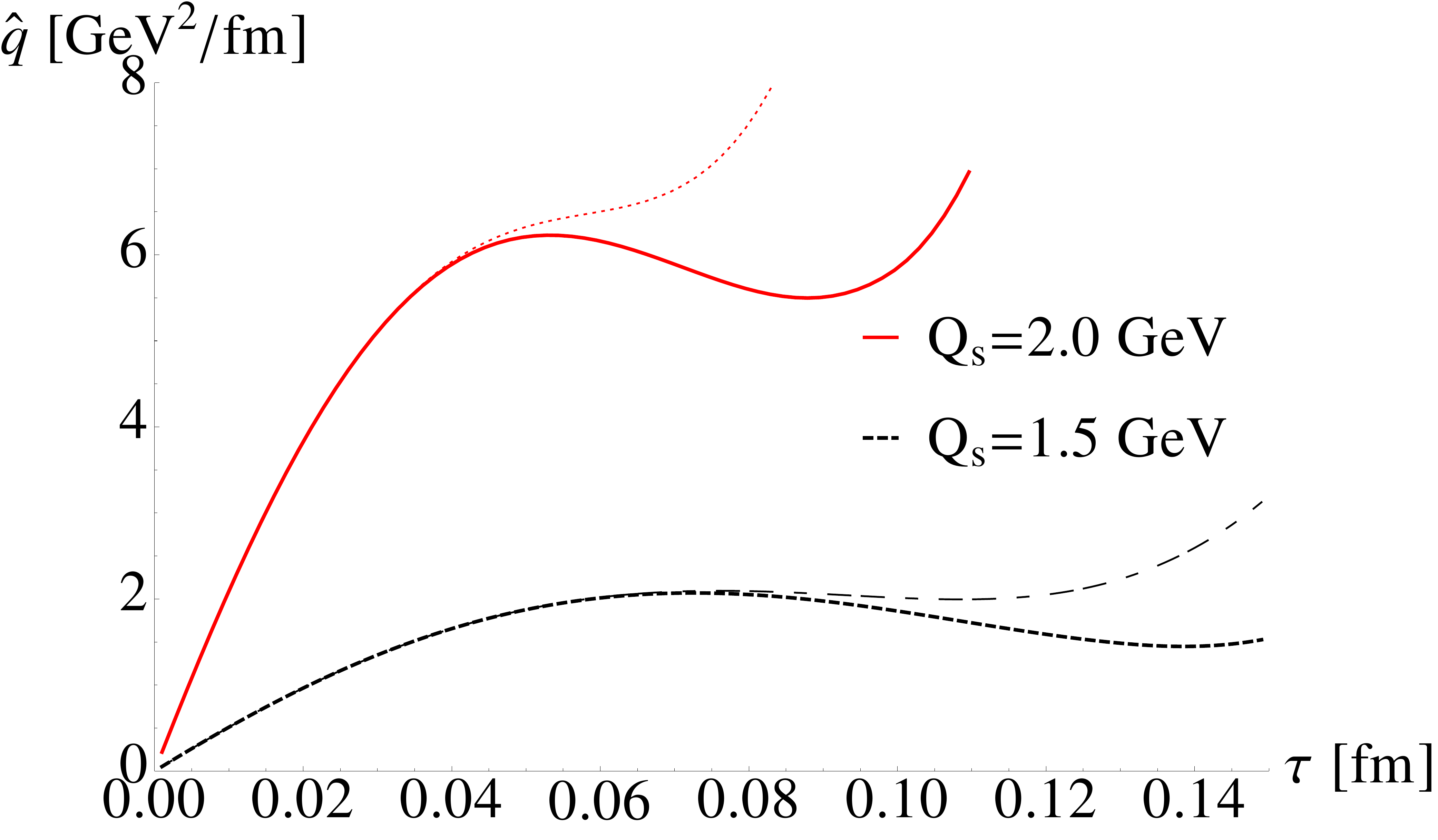}    
\includegraphics[scale=0.19]{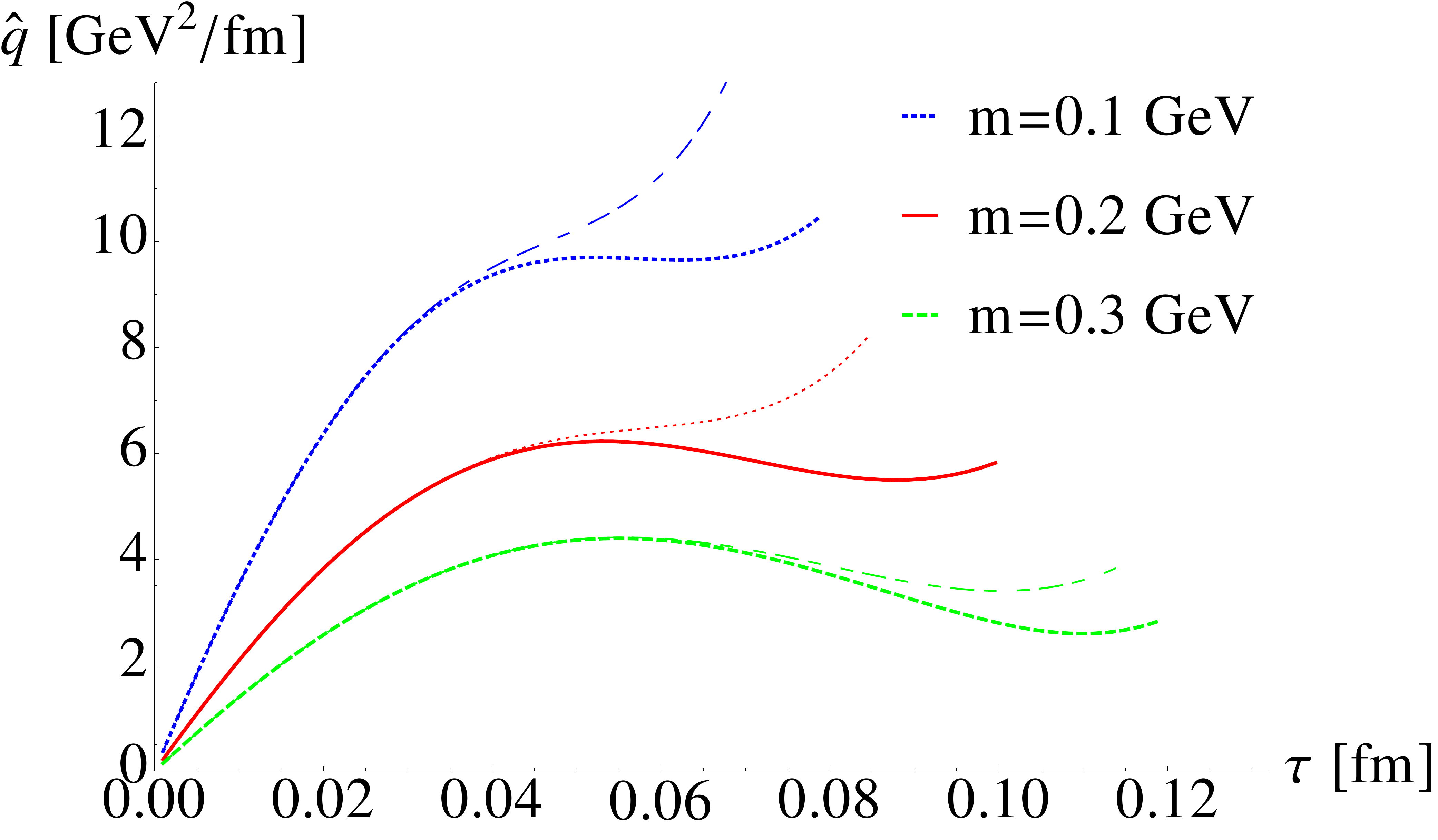}     
\caption{Left panel: The time evolution of $\hat q$ for two different values of $Q_s$ with $m=0.2$ GeV. Right panel: The time evolution of $\hat q$ for three different values of the infrared regulator $m$ with $Q_s=2$~GeV.
\label{param}} 
\end{figure}

The momentum broadening coefficient for different values of $Q_s$ and $m$ are depicted in Fig.~\ref{param} where, as previously, orders $\tau^4$ and $\tau^5$  are shown. Throughout this work we have been using $Q_s=2$ GeV and $m=0.2$ GeV, which is shown as the solid (red) lines in both panels. We observe that by decreasing $Q_s$ or increasing $m$ one can get smaller values of $\hat q$, and the $\tau$ expansion is reliable to longer times. 
From the left panel one finds that the results are reliable to about $\tau\sim 0.13$ fm for $Q_s=1.5$ GeV and $m=0.2$ GeV, and from the right panel this time is approximately $\tau \sim 0.11$ fm for $Q_s=2$ GeV and $m=0.3$ GeV.

To interpret these results we note that physically it makes sense to treat the saturation scale as a scaling parameter for the collision energy.  In collisions at RHIC, saturation is achieved at a scale $Q_s\sim 1-2$~GeV, as compared to higher energy collisions at the LHC, for which $Q_s \sim 2-3$~GeV \cite{Iancu:2003xm}. Our calculation thus predicts the dependence of $\hat q$ on the collision energy.  The effect  has in fact been seen in measurements of $\hat q$ produced during the later hydrodynamic phase. A reduction in $\hat q$ at  RHIC energies when compared to  LHC energies was found by the JET Collaboration in Ref.~\cite{JET:2013cls}, see also Ref.~\cite{Andres:2016iys}. 

We comment that the sensitivity of $\hat q$ to the scales $Q_s$ and $m$ is not unexpected, because the narrowness of the allowed momentum range makes it inevitable that shifting its upper and lower limits will affect results. 
In figure 9 we show the dependence of $\hat q$ on $Q_s$ with the ratio $Q_s/m$ held fixed. The figure shows that when the ratio of the two scales is constant, the dependence of the momentum broadening parameter is fairly weak. 

\begin{figure}[h]
\centering 
\includegraphics[scale=0.25]{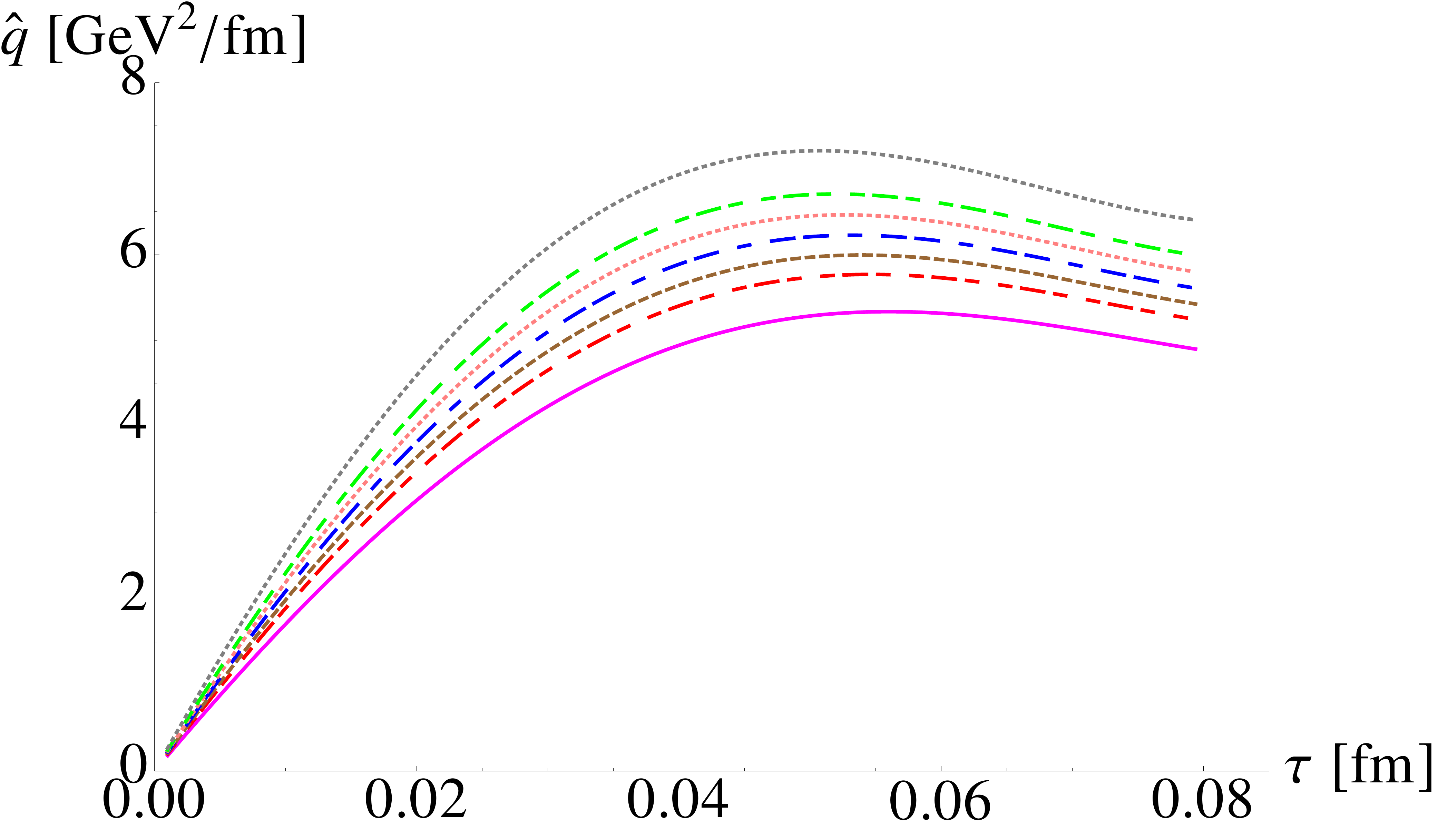}        
\caption{The momentum broadening coefficient $\hat q$ versus $\tau$ for different values of $Q_s$ with $Q_s/m=10$. The curves are labelled with different colours and patterns from $Q_s=1.9$ GeV (magenta solid line) to $Q_s=2.1$ GeV (grey dotted line).
\label{param-p}} 
\end{figure} 

\subsection{Regularization dependence}
\label{sec-res-c}

All results presented in this work have been obtained by introducing a regularization procedure. 
This regularization is needed because the function $C_1(r)$ defined by Eq.~(\ref{C1-def}) is divergent in the limit $r \to 0$. The correlators that enter the tensor $X^{\alpha\beta}({\bf v})$ in Eq.~(\ref{tensor-gen}) are determined by the functions $C_1(r)$ and $C_2(r)$, and their derivatives, and the lower limit of the integral over $t'$ corresponds to the point where $r=0$. We note that this divergence is a natural consequence of the fact that the CGC approach breaks down at small distances. 

To check that our results are largely independent of the regularization, we use two different methods to regularize the divergence, and compare the results. 
To explain this, we write the integrand for either momentum broadening or collisional energy loss as a function of the form $f(t',r,z)$, so that the transport coefficient is obtained from the integral $\int_0^t dt'\, f(t',r,z)$  (see Eqs.~(\ref{e-loss-X-Y}), (\ref{qhat-X-T}) and (\ref{tensor-gen})).
The first regularization method we use is to cut off the singular part of the integrand at a distance $r_s=Q_s^{-1}$ by defining the regularized function 
\be
\label{regularization1}
f^{\rm reg. 1}(t',r,z) \equiv \Theta(r_s-r)  f(t',r_s,z) + \Theta(r-r_s) f(t',r,z) .
\ee
We then obtain the transport coefficient from the integral $\int_0^t dt' f^{\rm reg. 1}(t',v_\perp t',v_\parallel t')$.
This method of regularization was used in all results presented above, as well as in our previous computations in Ref.~\cite{Carrington:2020sww}. 
The second regularization method is to subtract the leading order $\mathcal{O}(1/r)$ divergences before multiplying by the step function. We can represent this by defining $\tilde f (t',r,z)= f(t',r,z) -a/r$ with $a=\lim_{r\to 0} rf(t',r,z)$ and writing
\be
\label{regularization2}
f^{\rm reg. 2}(t',r,z) \equiv \Theta(r_s-r)  \tilde f(t',r_s,z) + \Theta(r-r_s) f(t',r,z) .
\ee 
The transport coefficient is obtained from the integral $\int_0^t dt' f^{\rm reg. 2}(t',v_\perp t',v_\parallel t')$. 

\begin{figure}[h] 
\centering
\hspace*{-0.6cm} \quad 
\includegraphics[scale=0.19]{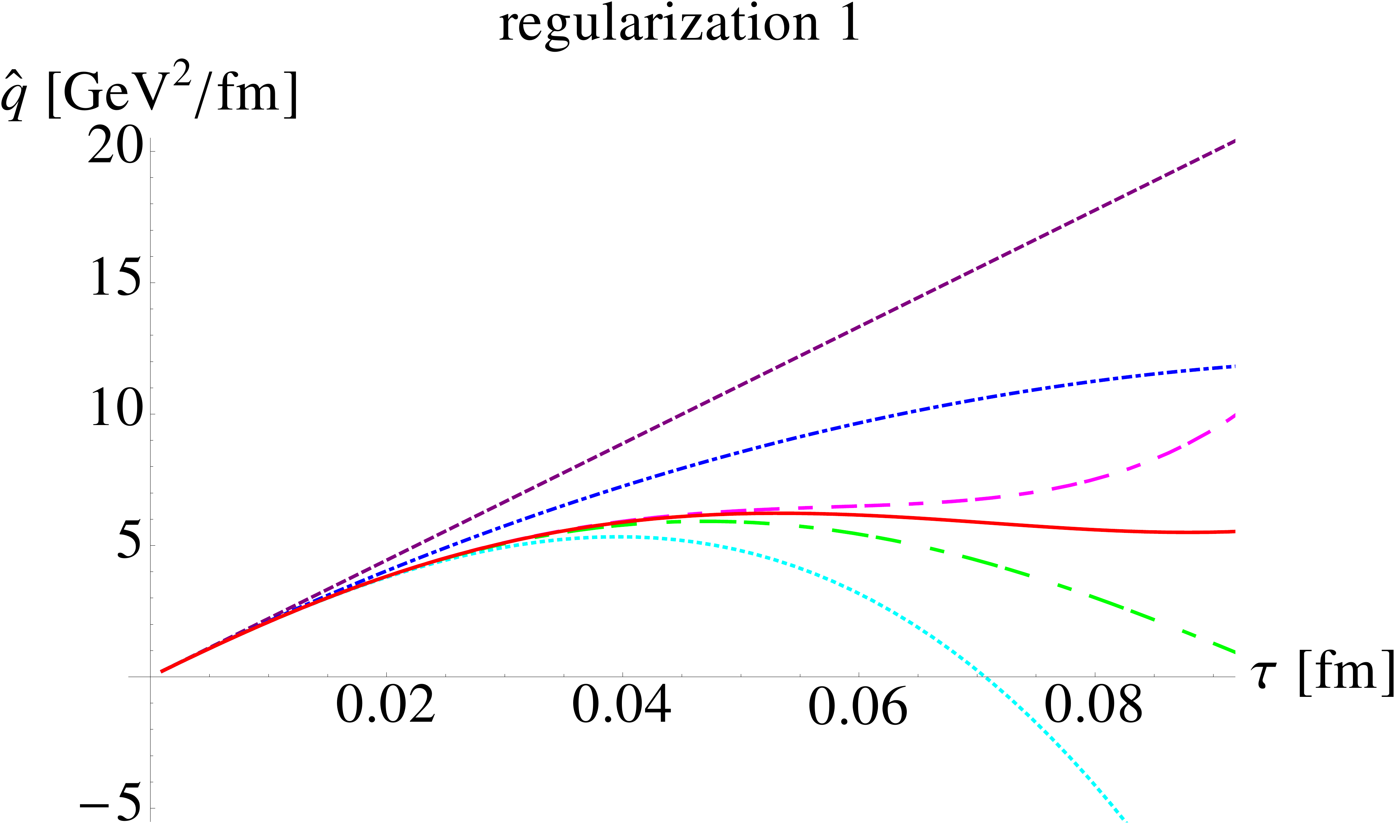}  
\includegraphics[scale=0.19]{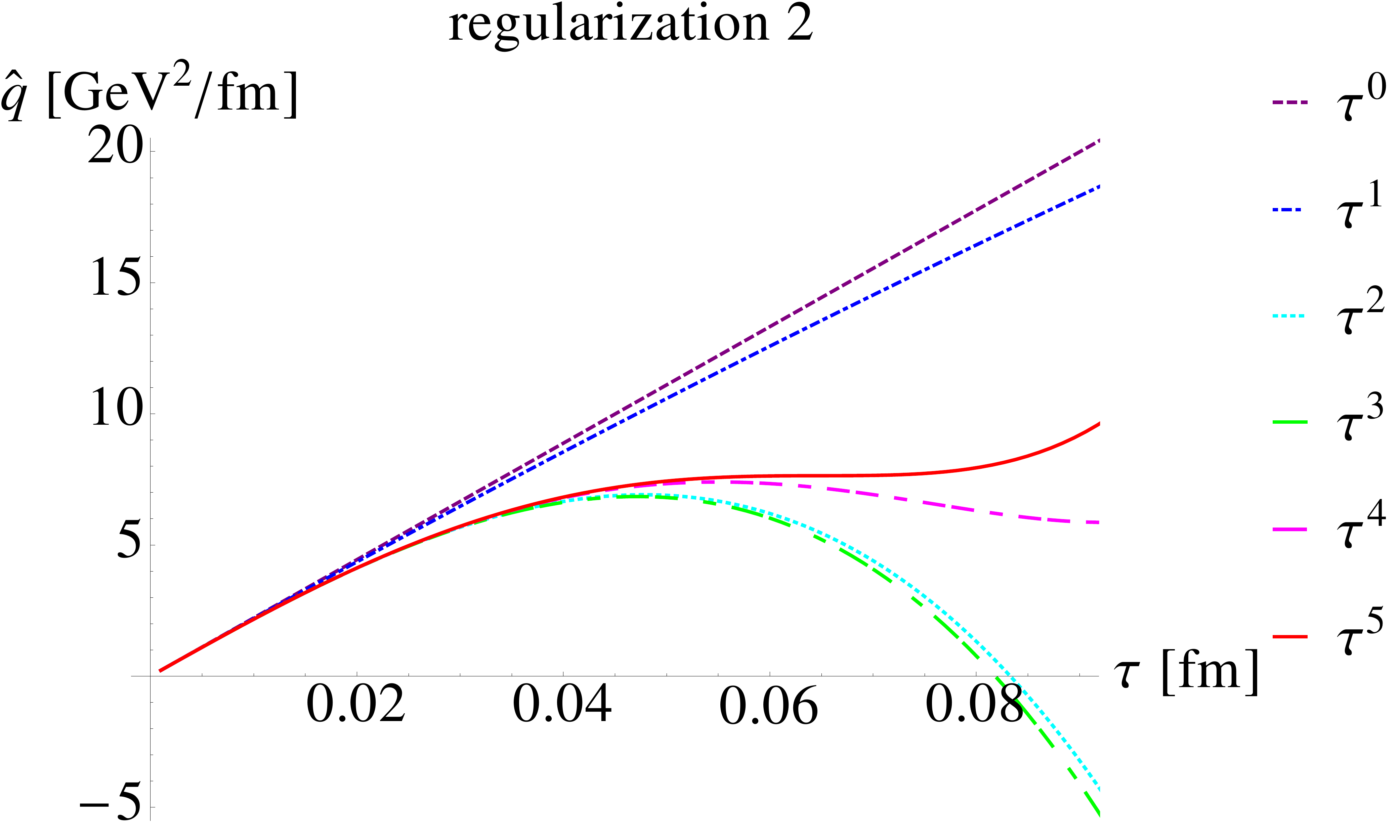}  
\caption{Cumulative results for $\hat q$ when higher and higher terms in the $\tau$ expansion are included using $v=v_\perp=1$.
\label{reg-sum} }
\end{figure}

\begin{figure}[h]
\centering 
\includegraphics[scale=0.25]{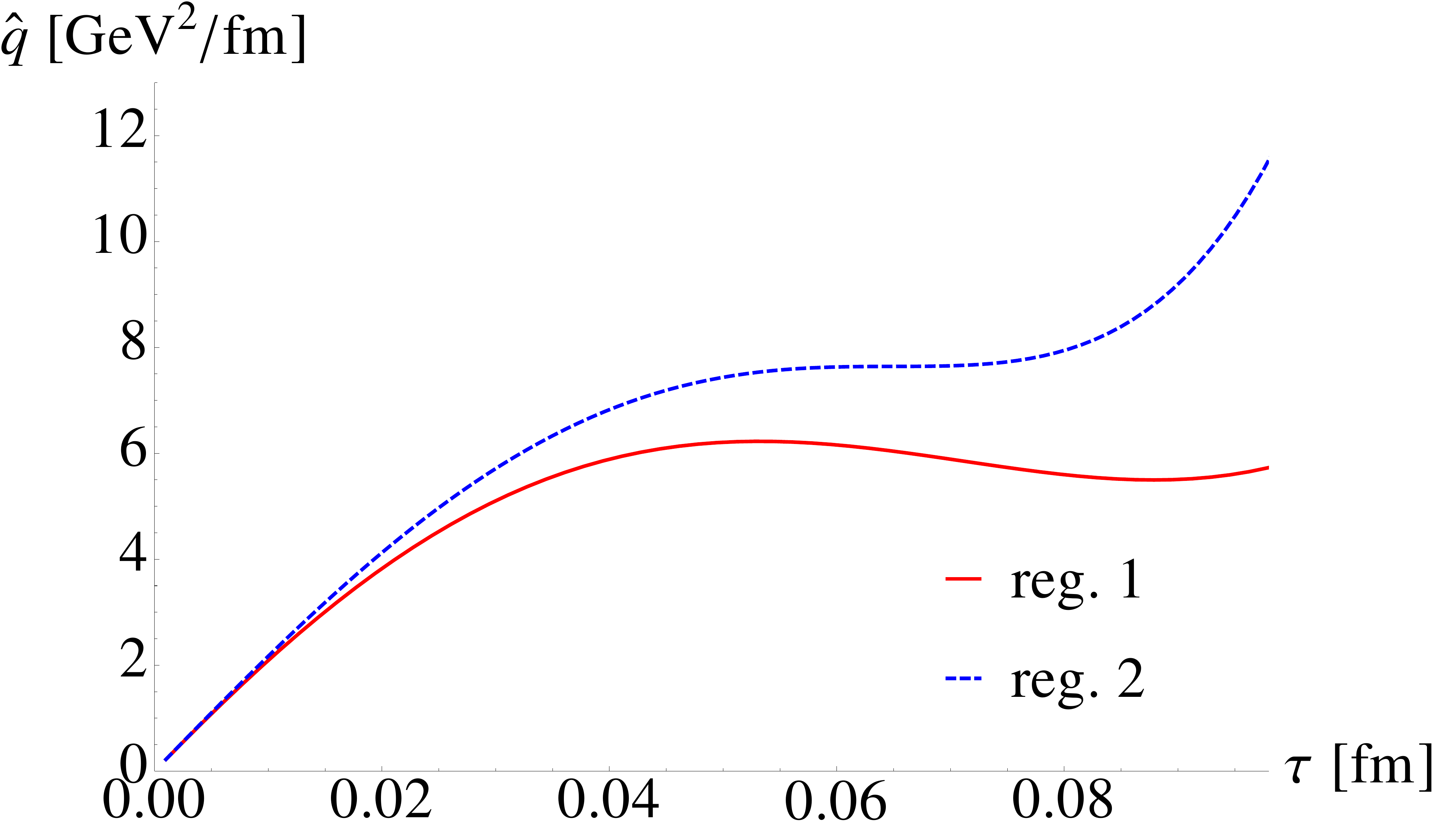}    
\caption{The time evolution of $\hat q$ at order $\tau^5$ for $v=v_\perp=1$ regularized in two different ways (see Eqs.~(\ref{regularization1}) and (\ref{regularization2})). 
\label{reg-sum-plot}}
\end{figure}

The results of the two different regularization methods are depicted in the left and right panels of Fig.~\ref{reg-sum}, and  comparison shows that the dependence on the regularization is fairly weak. In Fig.~\ref{reg-sum-plot} we show the fifth order results on the same graph.

\subsection{Limitations of the formulation}
\label{sec-res-e}


Our results, for either transport coefficient, can be trusted if two conditions are satisfied: if
there is evidence of saturation, which means that the Fokker-Planck approach we are using
is valid, and if the $\tau$ expansion converges, which means that our $\tau$ expanded glasma
correlators are reliable. In some cases, as in the left panel of Fig.~\ref{t-coeff-cum}, both of these
approximations break down at approximately the same value of $\tau$. This indicates that the
regime where our calculation works could be extended by working at higher order in the $\tau$
expansion. If the saturation region does not appear before the $\tau$ expansion fails, as in Fig.~\ref{two-cases}, the result cannot be trusted. 
It is also possible to encounter the opposite situation, where saturation occurs in a regime where the proper time expansion is not valid. In the calculations presented in our previous paper \cite{Carrington:2020sww}, where we worked to first order in $\tau$, 
saturation was observed for both transport coefficients at $\tau \approx 0.7$ fm. 
However the order $\tau^5$ results presented in this paper reveal that the proper time expansion breaks down at much earlier times. 
We emphasize that both saturation and the convergence of the proper time expansion are required. 

In Sec. \ref{sec-res-b} we explained why our approach works most efficiently for probes that move mostly transversely. 
Our method is therefore best suited to study the momentum broadening of a probe with $v_\perp$ close or equal to $v$, and works less well when $v_\parallel$ is large. 
We remind the reader that, for the most part, only small values of $v_\parallel$ are experimentally interesting, and collisional energy loss is always small in this region. 

Our results show that the transport coefficients that we calculated at order $\tau^5$ can be trusted to 
approximately $\tau\sim 0.05-0.13$ fm, depending on the choice of several parameters: the velocity of the probe, the saturation scale, and the value of the infra-red regulator.
It is interesting to compare this radius of convergence 
with the results we obtained in our previous papers, where we also used a proper time expansion.
In Refs.~\cite{Carrington:2020ssh,Carrington:2021qvi} we calculated the energy-momentum tensor, and obtained from it many physical quantities, including the energy density, transverse and longitudinal pressures, radial flow, several different measures of glasma anisotropy, Fourier coefficients of the azimuthal flow, and the angular momentum of the glasma. These quantities were calculated at order $\tau^5$ or $\tau^6$ (in most of these calculations only odd or even powers of $\tau$ contributed),
and we consistently used $Q_s=2$~GeV and $m=200$ MeV. 
One significant difference is that although the energy momentum tensor 
is constructed from the same 2-point correlator  as in this paper (see Eq.~(\ref{B-res})), the calculation requires that $r$ is taken to zero.
Therefore, the regularization was performed by imposing a cutoff on all momentum integrals at the saturation scale $Q_s$ and taking the limit that the spatial coordinate $r$ goes strictly to zero.
The behaviour of the $\tau$ expansion was not exactly the same for all of the quantities calculated in Ref.~\cite{Carrington:2021qvi}, as expected, but in
all cases the radius of convergence was $\tau \approx 0.05$~fm. This finding is comparable with the result found in this paper for the momentum broadening coefficient.

Finally we mention that our method is based on a CGC approach that is classical, and
there is an inherent lower bound on the proper time below we can no longer trust a classical
description. This bound can be estimated using the uncertainty principle. The very
large initial energy released in the collision produces a lower bound for the validity of
the classical description that is orders of magnitude smaller than the radius of convergence
of the $\tau$ expansion \cite{Carrington:2020ssh}.

\section{Glasma impact on jet quenching}
\label{sec-res-add}

We have found that in the glasma phase the momentum broadening parameter $\hat q$ can be as large as $\hat{q} \approx 6 ~ {\rm GeV^2/fm}$. 
The value of $\hat{q}$ in equilibrium quark-gluon plasma for a hard quark of $p_T > 40$ GeV is $2 < \hat{q}/T^3 < 4$ where $T$ is the plasma temperature, as inferred from experimental data by the JETSCAPE Collaboration \cite{JETSCAPE:2021ehl}. 
In the discussion below we take $\hat{q} = 3T^3$. 
Since the temperature of the plasma produced at the LHC evolves from roughly 450 to 150 MeV \cite{Shen:2011eg}, the momentum broadening coefficient varies from $\hat{q} \approx 1.0~{\rm GeV^2/fm}$ to $\hat{q} \approx 0.05~{\rm GeV^2/fm}$, which is much smaller than the value $\hat{q} \approx 6 ~ {\rm GeV^2/fm}$ for the glasma that we have obtained in this work. 
However, since the pre-equilibrium phase exists for less than 1 fm, 
it is not clear if the glasma
contributes significantly to the total momentum broadening that the probe experiences when it moves through the system.

\begin{figure}[h]
\centering
\includegraphics[scale=0.3]{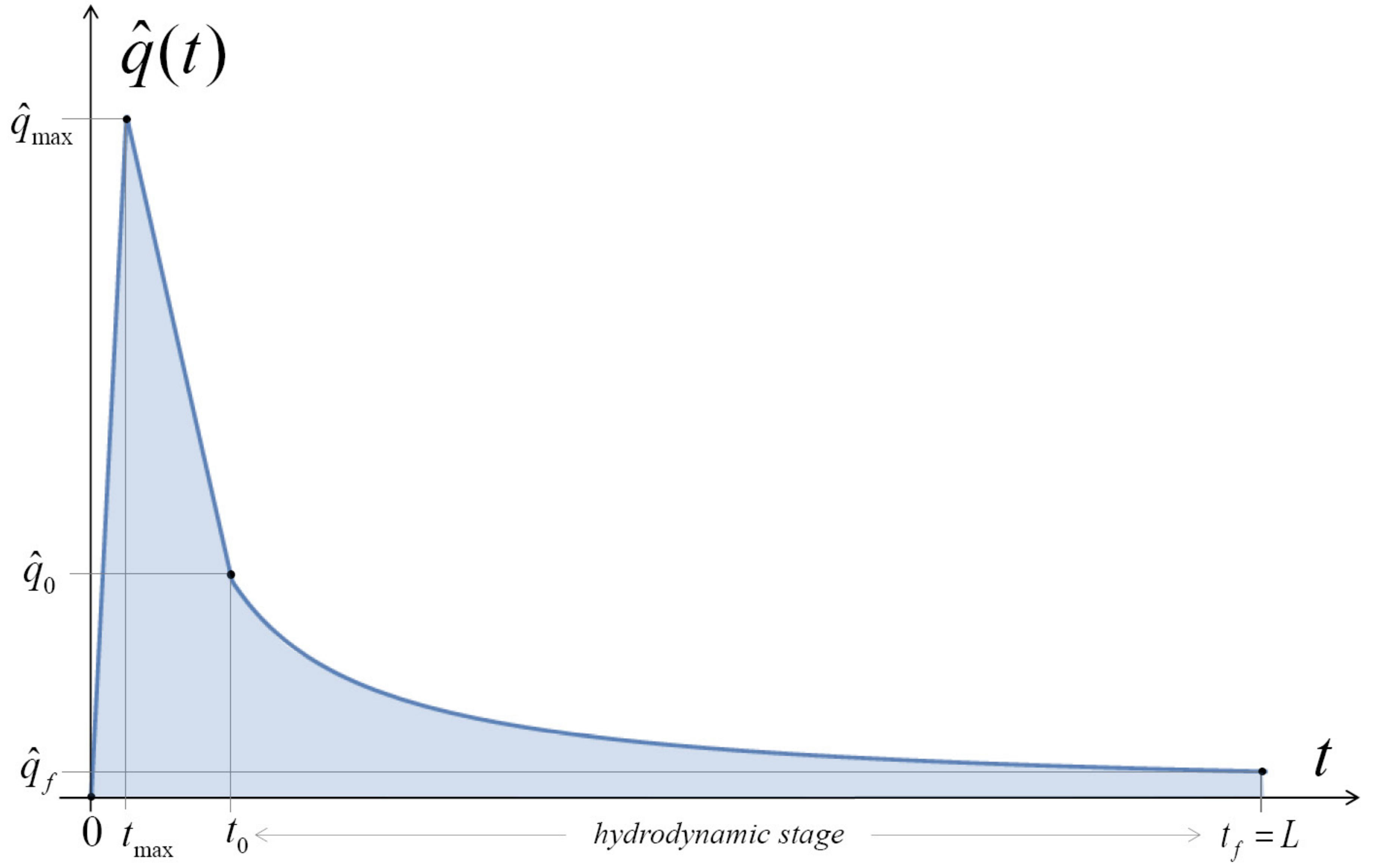}
\caption{Schematic representation of the temporal evolution of $\hat{q}(t)$. } 
\label{Fig-qhat-time}
\end{figure}

The radiative energy loss per unit length of a high-energy parton traversing a medium of length $L$ is proportional to the total accumulated transverse momentum broadening, denoted  $\Delta p_T^2$.
In case of a static medium, where $\hat q$ is constant, we have $\Delta p_T^2 = \hat{q}L$. When the plasma is not static and $\hat{q}$ is time dependent, the transverse momentum broadening is 
\be
\label{total-broadening}
\Delta p_T^2 = \int_0^L dt \, \hat{q}(t),
\ee
where the probe is assumed to move with the speed of light.  

Figure~\ref{Fig-qhat-time} is a schematic representation of the time dependence of the momentum broadening coefficient throughout the whole history of the probe's journey across the deconfined matter produced in a relativistic heavy-ion collision. The first part of the figure shows the rapid growth of $\hat{q}(t)$ to a maximal value $\hat{q}_{\rm max} \approx 6~{\rm GeV^2/fm}$ at $t_{\rm max} \approx 0.06$ fm. This is a rough description of the evolution of $\hat q$ in the glasma phase that we have found by working at order $\tau^5$. 
The value of $\hat q(t)$ subsequently decreases. At $t_0 \approx 0.6$ fm it has the value $\hat{q}_0 \approx 1.4$~GeV$^2$/fm 
(these numbers are estimates inferred from experimental data and are discussed in more detail below). 
We comment that the saturation region observed clearly in the left panel of Fig.~\ref{t-coeff-cum}
is not seen in Fig.~\ref{Fig-qhat-time}
because different
time scales are used in the two figures. 
The time interval between
$t_{\rm max}$ and $t_0$ is beyond the region of validity of the proper time expansion and the rapid decrease of $\hat q$ in this domain is not captured by our calculation, but it is
reproduced by the simulations in Ref.~\cite{Ipp:2020nfu}. Using linear interpolation between the points $\hat{q}(0)=0$, $\hat{q}(t_{\rm max})=\hat{q}_{\rm max}$, and $\hat{q}(t_0)=\hat{q}_0$, one finds the following non-equilibrium contribution to the accumulated transverse momentum broadening
\be
\label{non-eq}
\Delta p_T^2\Big|^{\rm non-eq} =  \int_0^{t_0} dt \, \hat{q}(t) 
= \frac{1}{2}\hat{q}_{\rm max} t_0 +  \frac{1}{2}\hat{q}_0(t_0 - t_{\rm max}) .
\ee

At $t>t_0$ we have equilibrated quark-gluon plasma which expands hydrodynamically. 
Using  ideal one-dimensional boost invariant hydrodynamics the temperature decreases as
\be
T = T_0 \Big(\frac{t_0}{t}\Big)^{1/3}. 
\ee
Consequently, the momentum broadening coefficient depends on time as
\be
\hat{q}(t)  = 3 T_0^3 \, \frac{t_0}{t} = \hat{q}_0 \frac{t_0}{t}
\ee
and the equilibrium contribution to $\Delta p_T^2$ is
\be
\label{eq}
\Delta p_T^2\Big|^{\rm eq} =  \int_{t_0}^L dt \, \hat{q}(t) 
= 3 T_0^3 \, t_0 \, \ln \frac{L}{t_0} .
\ee

To estimate the role of the glasma in jet quenching we need, in addition to $\hat{q}_{\rm max}$ and $t_{\rm max}$ which come from our calculation, the following parameters: $T_0$, $t_0$, $\hat q_0$ and $L$. The time $t_0$, which marks the beginning of the hydrodynamic evolution, and the initial temperature $T_0$, which determines the system's initial energy density, are obtained by comparing hydrodynamic models with experimental data on particle spectra and collective flows. 
The initial time cannot be too small as the system should reach, at least approximately, local thermodynamic equilibrium 
for a hydrodynamic approach to be applicable. On the other hand the initial time cannot be too big, because in that case the initial shape of the system would be washed out and hydrodynamics would not be able to reproduce the Fourier coefficients of the collective flow. We use $T_0 = 0.45$ GeV and $t_0 = 0.6$~fm taken from \cite{Shen:2011eg} and \cite{JETSCAPE:2021ehl}, respectively. The momentum broadening parameter is inferred from experimental data on jet quenching through complex modelling of the process of hard probe propagation through the evolving plasma. Using again the results of the JETSCAPE Collaboration \cite{JETSCAPE:2021ehl}, we take $\hat q_0 \approx 3 T_0^3 \approx 1.4$ GeV$^2$/fm. Finally, keeping in mind that the radius of a heavy nucleus (for example, Au or Pb) is about 7 fm, we assume that the typical path length of a hard probe in the quark-gluon plasma is $L = 10$ fm. 
The length scale $L$ is chosen to be slightly bigger than a typical nuclear radius, because the effect of jet quenching is particularly evident when the point of the jet production is close to the system's surface. 
In this case, one jet easily escapes into vacuum while the jet going in the opposite direction propagates through the plasma and, in central collisions, its path can be as long as the diameter of the nucleus. 
Substituting these values into Eqs.~(\ref{non-eq}) and (\ref{eq}), we find
\be
\label{ratio}
\frac{\Delta p_T^2\Big|^{\rm non-eq}}{\Delta p_T^2\Big|^{\rm eq}} = 0.93.
\ee
We note that this result is not very sensitive to the  parameters $T_0$, $t_0$, $\hat q_0$ and $L$, or the precise shape of the peak in Fig. \ref{Fig-qhat-time}. 
Equation (\ref{ratio}) shows that the non-equilibrium phase gives a contribution to the radiative energy loss which is comparable to that of the equilibrium phase. 
The conclusion is that the glasma plays an important role in  jet quenching, which contradicts the commonly made assumption that the contribution of the glasma phase to momentum broadening is negligible.

\section{Summary, discussion and conclusions}
\label{sec-conclusions}


In this paper we have calculated the collisional energy loss and momentum broadening of hard probes moving through the strongly interacting matter from the earliest phase of a heavy-ion collision. We have combined two approaches. The medium that the hard probe interacts with is a glasma described in terms of a CGC effective theory with a proper time expansion. This description applies only at very early times. We use a Fokker-Planck equation to describe the interactions of the hard probe with the chromodynamic fields populating the glasma. A Fokker-Planck description is valid only at sufficiently long times that saturation of the collision terms occurs. Therefore, there is an inherent conflict between the assumptions that set the time scales for the two parts of our calculational method. In addition, a Fokker-Planck description requires gradient expansion type approximations, and the CGC approach that we use assumes boost invariance (see sections \ref{FP-eq} and \ref{CGC-all} for details). It is not {\it a priori} clear that all of these different conditions can be satisfied simultaneously.

Our calculation allows us to directly verify the validity of our method by comparing the range of proper times for which the Fokker-Planck collision integrals saturate, and the proper time expansion converges. 
In the region of space-time between the two ions, post-collision, the glasma fields extend widely across the transverse plane, but the regions over which these fields are correlated are much smaller (see Fig.~\ref{tubes}). 
This structure provides a mechanism for the saturation of the time dependence of $\hat q$ and $dE/dx$, which occurs when a probe leaves the region of the glasma where highly correlated fields exist. 
The extent of the region of correlated fields is determined by the form of the $\tau$ expanded CGC correlators. 
If the velocity of the probe allows it to escape from the region of correlation before the $\tau$ expansion breaks down, the transport coefficients determined from the Fokker-Planck equation can saturate. 
When this behaviour is observed, it indicates that the approximations introduced in the two different components of our method are simultaneously satisfied.

We have shown that in many cases saturation occurs within the radius of convergence of the proper time expansion. 
The key variable is the orientation of the probe's velocity. 
Saturation occurs earlier when a probe moves mostly perpendicularly to the beam axis. 
This means that the domain where our approach works best coincides with the momentum space mid-rapidity region $y \approx 0$ where jet quenching is studied experimentally. 
Our method can be used to obtain a reliable estimate of $\hat q$ with $v_\perp$ close to $v$. 
Collisional energy loss is much more difficult to calculate using our method. 
Saturation is not really observed, and our calculations can give at best order of magnitude estimates. 
However, collisional energy loss is also less important because it is small in the kinematic region of small or vanishing $v_\parallel$. 

The momentum broadening coefficient $\hat q$ saturates at $\tau\approx 0.07$ fm for $v=v_\perp=1$. 
Its value depends somewhat on the values of the saturation momentum $Q_s$ and infra-red regulator $m$: it grows with increasing $Q_s$ at fixed $m$, and decreases with increasing $m$ at fixed $Q_s$. 
Using the typical values $Q_s=2$ GeV and $m=0.2$ GeV we find that $\hat q$ obtains a maximal value of approximately $6$ GeV$^2$/fm. 
The value of $\hat q$ is only weakly dependent on the spatial rapidity $\eta$, and the procedure  employed to regulate the Fokker-Planck collision integrals. 
We have also calculated the total accumulated momentum broadening and shown that the glasma phase gives a contribution that is comparable to that of the long lasting equilibrium phase. This result indicates that the standard practice of ignoring the contribution of the glasma phase to momentum broadening is unjustified. 

It is important to consider whether the systems produced at RHIC and the LHC would be accurately described by the method we use, which makes use of several simplifying approximations. 
We apply the MV model, and a proper time expansion, which means that we assume that the collision can be described in terms of two infinitesimally thin nuclei with infinite extent in the transverse plane,
that collide at $\tau=0$ fm.
The glasma is produced at the moment of the collision, and then evolves in time. 
This is simplified picture does not take into account the finite time required for the nuclei to pass through each other, due to their finite width. 
This time will have an effect on the dynamics of the system, and the effect will be larger at lower collision energies. A theoretical approach to include these effects has not been developed. 
At RHIC energies the time for two nuclei to pass through each other is of order 0.1 fm. 
At first glance this appears very troubling, since the radius of convergence of the expansion we are using is approximately $0.05-0.07$~fm. However, as explained above, our calculation assumes that the glasma is formed at $\tau=0$. 
In reality, the glasma is not formed when the nuclei initially make contact, 
and therefore it presumably does not make sense to include the full duration of the time required for them to pass through each other within the interval where the proper time expansion is valid. 
In other words, the initial time in our calculation might reasonably be taken to correspond to a time somewhere between the point of initial contact, and the time the nuclei have passed through each other. 
In any case, at  LHC energies the time required for two nuclei to pass through each other is an order of magnitude smaller, and thus well within the radius of convergence of the proper time expansion. 
We also point out that in this work we are primarily interested in hard probes that propagate mostly in the transverse direction, which are most relevant experimentally. The physics of high $p_T$ probes is presumably not strongly affected by longitudinal dynamics that is not correctly taken into account due to the assumption of vanishing widths of colliding nuclei.  

Let us comment on the possible role of Weibel instabilities in the evolution of the glasma. 
Weibel instabilities appear in an anisotropic system when there is a transmission of energy from the plasma constituents to fields, see the review \cite{Mrowczynski:2016etf}.  
In the approach we have used, quasi-particles are not present and there are only soft classical fields, so there is no mechanism that could generate plasma instabilities.
We note that in Refs.~\cite{Romatschke:2006nk,Romatschke:2005pm} it has been proposed that the role of particles could be played by the hard modes of the glasma, and unstable modes are found as solutions
to the Yang-Mills equation. However, these calculations require that boost invariance is broken, and the effect cannot be seen in our boost invariant formalism.

Finally, we comment that 
it would be very interesting to study the dependence of collisional energy loss and momentum broadening on the impact parameter of the collision. 
This calculation would provide verification of the dependence of jet quenching on the collision centrality. 
In our previous paper \cite{Carrington:2021qvi} we have developed a method to use a Woods-Saxon distribution for the nuclear density, and calculate field correlators using a gradient expansion of the charge density. 
The analysis is technically involved and is left for future work.

\section*{Acknowledgments}

This work was partially supported by the National Science Centre, Poland under grant 2018/29/B/ST2/00646, and by the Natural Sciences and Engineering Research Council of Canada under grant SAPIN-2017-00028.


\end{document}